# Measurement of Nanoscale Surface Disturbances on Liquid Jets Using the Mie Scattering Method

Dingwei Zhang[1], Qiyou Liu[1], Hu Sun[1], Qingfei Fu[1,2,3], Bingqiang Ji[1,2,4*], Lijun Yang[1*]

[1] School of Astronautics, Beihang University, Beijing 100191, PR China

[2] Aircraft and Propulsion Laboratory, Ningbo Institute of Technology, Beihang University, Ningbo 315800, PR China

[3] National Key Laboratory of Aerospace Liquid Propulsion, Xi'an, 710100, PR China

[4] State Key Laboratory of High-Efficiency Reusable Aerospace Transportation Technology, Beijing 100191, PR China

**Corresponding authors:**

*bingqiangji@buaa.edu.cn (B.J.); *yanglijun@buaa.edu.cn (L.Y.)

**Abstract**

The Plateau-Rayleigh instability of liquid jets governs a wide range of natural and industrial processes. While the classical linear stability theory accurately predicts disturbance growth rates and dominant wavelengths, it cannot quantify the amplitude and frequency spectrum of nanoscale initial surface disturbances. This critical missing information prevents accurate forecasting of breakup lengths and droplet generation. Conventional optical imaging is constrained by the diffraction limit and lacks the capability to resolve such nanoscale surface fluctuations. Here we introduce a Mie-scattering-based measurement technique to characterize nanoscale surface disturbances on liquid jets. The method employs a thin laser sheet to illuminate the jet, with scattered light captured at designated azimuthal observation angles. Based on Lorenz–Mie theory, we develop an inversion algorithm incorporating cosine similarity and cross-correlation to reconstruct jet surface disturbances from measured scattering fringes, achieving a sub-nanometer resolution for detecting diameter variations, with an overall measurement precision limited by a background noise floor of 0.18 nm RMS. Comparative analyses of forward and rainbow scattering signals further confirm that the surface disturbances become axially symmetric as the amplitude reaches a few nanometers, and thus satisfies the assumptions of the scattering model. Synchronous validations against high-speed imaging and scanning electron microscopy corroborate the quantitative accuracy of our scattering method. Applied in the near-nozzle region, the technique identifies two intrinsic signatures of initial perturbations: stochastic pulse-like fluctuations and low-frequency oscillations. This work establishes a robust experimental platform for direct characterization of the initial conditions governing jet instability, paving a new pathway toward a complete mechanistic understanding of jet breakup from its nascent origins.

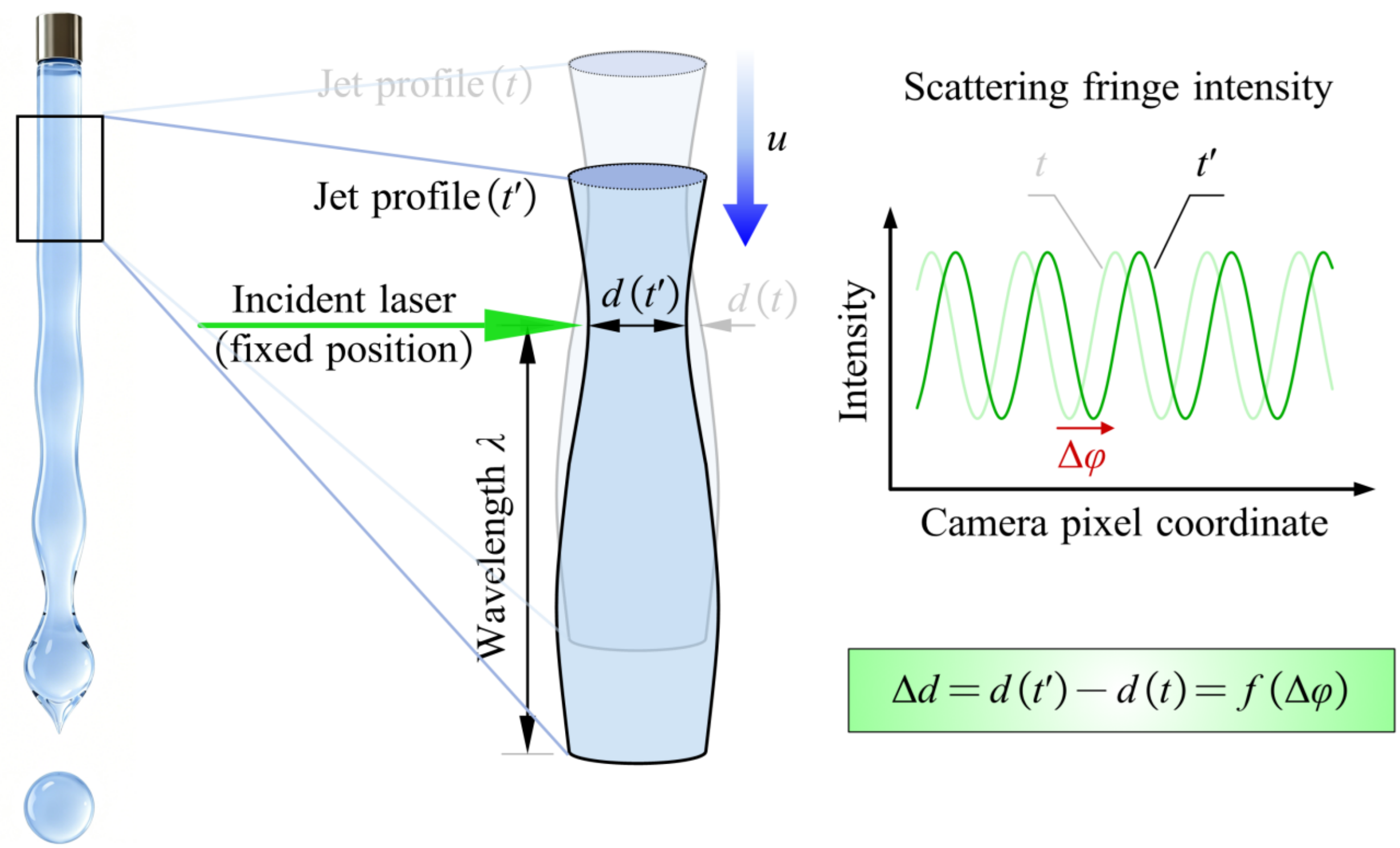

**Graphic abstract**

## 1 Introduction

As a representative example of interfacial instability, the breakup of a liquid jet into droplets is ubiquitous in both nature [1, 2] and industrial applications [3]. Spanning water-jet guided laser processing [4], inkjet printer [5], microfluidic devices in pharmaceuticals [6, 7], and tin droplet generators for EUV (Extreme Ultra-violet) lithography [8], the dynamics of jet instability governs the size, frequency and distribution of the generated droplets.

The seminal insight dates back to Plateau [9], who recognized in 1873 that surface energy minimization favours the breakup of a cylindrical jet into a periodic train of droplets. Rayleigh [10] placed this perception on quantitative basis by performing a linear stability analysis of an inviscid jet, demonstrating that axisymmetric disturbances of wavelength exceeding the jet circumference grow exponentially. Continuous growth leads the disturbance amplitude reach the jet initial radius $r_0$ at the downstream, causing the final breakup, where the most unstable mode occurs at a wavelength $\lambda \approx 9r_0$. Rayleigh's prediction, along with its subsequent and numerous extensions [11-15], have been confirmed with remarkable precision in terms of growth rate and wavelength [16-25]. Together, these contributions define the Plateau-Rayleigh instability that underpins our understanding of jet breakup.

Although the linear stability theory enables to obtain the growth rate $k_\mathrm{i}$ based on fluid properties (density $\rho$, viscosity $\mu$, surface tension $\sigma$), the jet radius and the frequency $f$ of the disturbance, it conspicuously fails to address a critical detail: the concrete form of the initial disturbance $\eta_0$, which is hypothesized to be on the nanoscale [21,24,25,26]. Without information regarding the amplitude and frequency spectrum of the initial disturbance, accurately predicting of jet breakup length, droplet formation frequency, and droplet size, remains an impractical task. Furthermore, for jets with diameters typically ranging from tens to hundreds of micrometers [16-26], the surface disturbances over nearly half the jet length remain at the submicron scale considering its exponential growth (Fig. 1). Thus, the characteristic of nanoscale disturbances serves as a key link in fully understanding the dynamics of jet breakup.

However, the nanoscale surface disturbances at the jet upstream are undetectable by conventional imaging techniques. In most previous studies, scholars could merely infer the initial disturbance amplitude indirectly according to the breakup length of the jet and the exponential growth rate [21, 24-26]. Most of the previous experiments measured the jet surface disturbances using a high-speed camera with a microscopic lens. However, constrained by a resolution limited the micrometer scale due to the well-known "diffraction limit" [27], this method only applied to disturbances of micron scale at the jet downstream. Existing methods for circumventing the optical diffraction limit, such as super-resolution imaging [28], electron microscopy [29] and X-ray imaging [30], are severely limited in liquid jet experiments owing to insufficient temporal resolution, the need for vacuum conditions or other constraints. Taub [18] proposed a laser occlusion method, which enhanced the resolution to the 0.1 μm level. However, this method necessitates the use of opaque liquids and strict calibration procedures. Therefore, it faces significant challenge in being widely adopted for general jet research.

Fortunately, characterizing liquid jet disturbances only requires the acquisition of jet diameter variations with time. This low data requirement makes it feasible to adopt non-imaging measurement techniques. When a coherent laser beam illuminates a small target, the resulting scattering interference fringes carry characteristic information about the size of the target. The scattering fringes can magnify the variation of the target size by hundreds or thousands times, which is known as the Mie scattering and can be well described by the Lorenz-Mie theory [31]. Mie scattering has become a mature technique in fluid research and is commonly used to measure spherical particles such as droplets and bubbles [32]. Furthermore, optical fibers, which share a cylindrical geometry with liquid jets, have achieved diameter measurements with nanometer-scale precision using Mie scattering methods [33, 34]. These existing applications verify the feasibility of employing Mie scattering for high-precision measurements of liquid jet disturbances. However, direct application of this technique to liquid jets is not trivial. First, unlike static optical fibers, a liquid jet is a dynamic medium with axially varying curvature which can cause severe distortion of the scattering fringes [35, 36], especially near the nozzle exit and downstream where the disturbance amplitudes exceed micrometer level. Second, environmental noises can introduce spurious small-amplitude fluctuations that are easily misidentified as genuine initial disturbances, thereby obscuring the true physical origins of jet instability at the nanoscale. Third, the inversion of measured fringe patterns into diameter variations is far from straightforward: the scattered light distribution changes with diameter in a highly intricate manner, making the mapping from a fringe vector to the corresponding diameter value a challenging task that demands a reliable and robust algorithm. Fourth, there exists no independent reference technique with sufficient precision to directly verify dynamic nanoscale disturbances on a jet, which inherently complicates the validation of the measurement of Mie scattering.

To address the above challenges, we put forward an optimized measurement method based on Mie scattering for measuring nanoscale surface disturbances on a circular liquid jet. The measurement principle, experimental setup, and inversion algorithm are described in Section 2. Section 3 addresses background noise characterization and noise suppression. Symmetry tests, along with validation against synchronous imaging and scanning electron microscopy are provided in Section 4. Section demonstrates the method's application in the near-nozzle region, revealing two intrinsic stochastic signatures of initial disturbances.

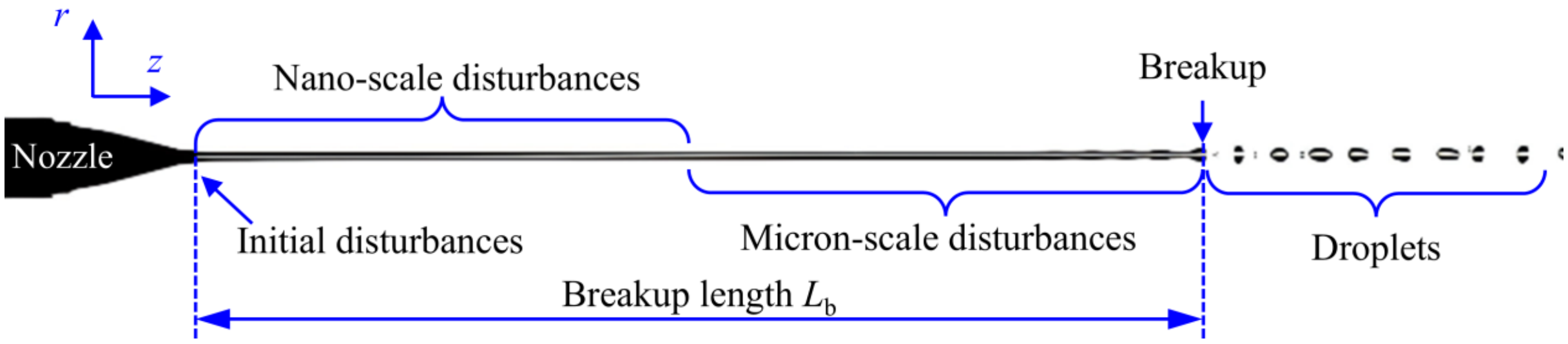

**Fig. 1** Schematic of the liquid jet breakup and the disturbance growth along the jet

## 2 Experimental methodology

### 2.1 Lorenz-Mie theory

The liquid jet is treated as an infinite, uniform dielectric cylinder, while the incident light is regarded as a plane electromagnetic wave, as depicted in Fig. 2a. The Mie scattering is thus described by the Lorenz-Mie theory [31], where the scattered electric field of the cylinder is expressed as:

$$E^{sca} = \left(\frac{2}{\pi k_E R}\right)^{1/2} \exp\left(-\mathrm{i}k_E R + \mathrm{i}\omega_E t - \mathrm{i}\frac{3\pi}{4}\right) T_1(\varphi) \tag{1}$$

$$T_1(\varphi) = \sum_{n=-\infty}^{\infty} a_n \exp(\mathrm{i}n\varphi) = a_0 + 2\sum_{n=1}^{\infty} a_n \cos(n\varphi) \tag{2}$$

where $k_E = 2\pi/\lambda$ with $\lambda$ being the light wavelength, $\omega_E = k_E c$ with $c$ represents the light speed, $R$ is the radius of the cylinder, $\varphi$ represents the scattering azimuth, and $a_n$ are scattering coefficients determined by refractive index $m_r$, $\lambda$ and $R$:

$$a_n|_{\mathrm{TM_z}} = -\frac{m_r J_n'(m_r k_E R) J_n(k_E R) - J_n(m_r k_E R) J_n'(k_E R)}{m_r J_n'(m_r k_E R) H_n^{(2)}(k_E R) - J_n(m_r k_E R) H_n^{(2)\prime}(k_E R)} \tag{3}$$

$$a_n|_{\mathrm{TE_z}} = -\frac{J_n'(m_r k_E R) J_n(k_E R) - m_r J_n(m_r k_E R) J_n'(k_E R)}{J_n'(m_r k_E R) H_n^{(2)}(k_E R) - m_r J_n(m_r k_E R) H_n^{(2)\prime}(k_E R)} \tag{4}$$

where $J_n$ denotes the *n*th order first kind Bessel function, $H_n^{(2)}$ stands for *n*th order second kind Hankel function, and $J_n'$ and $H_n^{(2)\prime}$ signify their respective derivatives. In the transverse magnetic ($TM_z$) polarization, the electric field of the light wave is oriented perpendicular to the jet. In the transverse electric ($TE_z$) polarization, the electric field is parallel to the jet. The light intensity is proportional to $|E^{sca}|^2$. Accordingly, the far-field scattered light intensity distribution can be expressed as:

$$I(\varphi) = I_0 \frac{2}{\pi L k_E} |T_1(\varphi)|^2 \tag{5}$$

where $I_0$ denotes the intensity of the incident light, and $L$ is the distance between the jet and the observation position. Fig. 2b depicts the intensity distribution of 532 nm $TM_z$ light scattered by a cylinder with a radius $R = 200$ μm and a refractive index $m_r = 1.33$. Owing to the relatively large diameter of the cylinder ($R/\lambda \gg 1$), the scattered light presents a complex distribution featuring dense oscillations along the angular direction. Therefore, it becomes essential to examine the sensitivity of the intensity distribution to variations in the jet diameter within a small angular range. Besides, a small angular range is precisely what can be effectively captured by a single camera. Because the scattered light under $TM_z$ illumination uniquely exhibits the typical "rainbow" pattern [36] at large scattering angle ($\varphi = 138°$), all subsequent experiments and discussion are based on $TM_z$ polarization. Forward scattering ($\varphi = 20°$) was also used, which offers lower sensitivity than rainbow scattering but is suitable for large-amplitude disturbances.

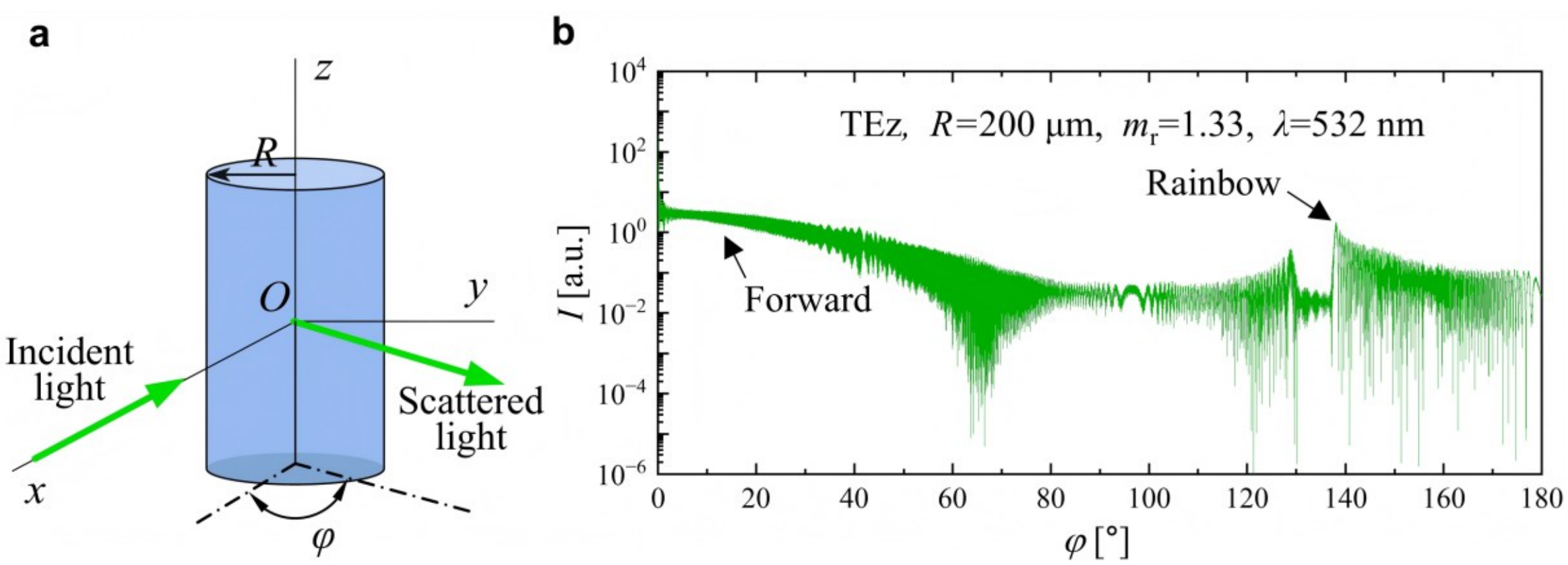

**Fig. 2 a,** Schematic of the Mie scattering generated by a cylinder**. b,** far-field light intensity distribution around a cylinder.

A rough estimate of the resolution for measuring cylindrical diameter variations based on Lorenz-Mie theory (see Supplementary Information and Fig. S1) gives a nanometer-level precision, demonstrating that Mie scattering is capable of resolving nanoscale disturbances in liquid jets.

## 2.2 Experimental setup

### 2.2.1 Jet test facility and optical setup

The experiments were conducted on a specially designed high-precision liquid jet test facility. Jets were generated by a gravity tank through a metal nozzle. Detailed information on the jet formation can be found in our previous work [26]. Although syringe pumps commonly used in previous jet experiments can precisely control the flow rate, they introduce considerable low-frequency noise, (see Supplementary Information and Fig. S3). The test media was deionized water at 25°C, with density $\rho = 997\ \mathrm{kg \cdot m^{-3}}$, viscosity $\mu = 0.9\ \mathrm{mPa \cdot s}$, surface tension $\sigma = 72\ \mathrm{mN \cdot m^{-1}}$, and refractive index $m_r = 1.33$. The nozzle used to generate the jet has an inner diameter of $d_0$ = 400 µm. For the thin water jet with a relative low speed ($u \leqslant 5$ m/s) in our experiments, the shear effect from the ambient gas and the gravitational stretching effect can be neglected, and thus the dynamics of the jet disturbance is governed by Weber number $We = \rho u^2 r_0/\sigma$ and Ohnesorge number $Oh = \mu/\sqrt{\rho\sigma r_0}$ [11, 37]. Thus, the jet radius is determined as the inner radius of the nozzle, i.e., $r_0 = d_0/2 = 0.2$ mm, and the jet velocity is calculated as $u = Q/(\rho\pi r_0^2)$ with $Q$ being the mass flow rate of the liquid.

Fig. 3a shows the schematic of the optical setup. A laser source (Lasertack, MXL-FN-532) emits a 532 nm beam with a diameter of 1.5 mm, which is then shaped by a beam expander, a slit, and a cylindrical lens into a thin laser sheet. At its thinnest point, located just beneath the nozzle exit, the sheet measures 7.5 mm in width and 28 μm in thickness. Upon interacting with the liquid jet, scattered light is generated and passes through a second cylindrical lens and a polariscope before entering microscope 1 (Leica Z16 APO) to be captured by camera 1 (Photron SA-Z). The distance $L$ between the focal plane of the camera and the jet, referred to as the off-focus distance, determines the number of fringes captured. A smaller off-focus distance yields denser scattering fringes in the camera image. In our experiments, this distance was kept at $L = 100$ mm.

Camera 2 (Photron Mini AX200), equipped with microscope 2 (Laowa Aurogon FF), is used to record a direct image of the jet. This imaging path relies on a 660 nm LED light source and a 532 nm cut-off notch filter to block the scattered laser light. Fig. 3b presents a photograph of the laser sheet used for measurement, showing the focused waist at the jet location. The sheet thickness gradually expands on both sides of this waist, while the sheet width remains unchanged, confirming the formation of a thin illumination plane precisely at the measurement region.

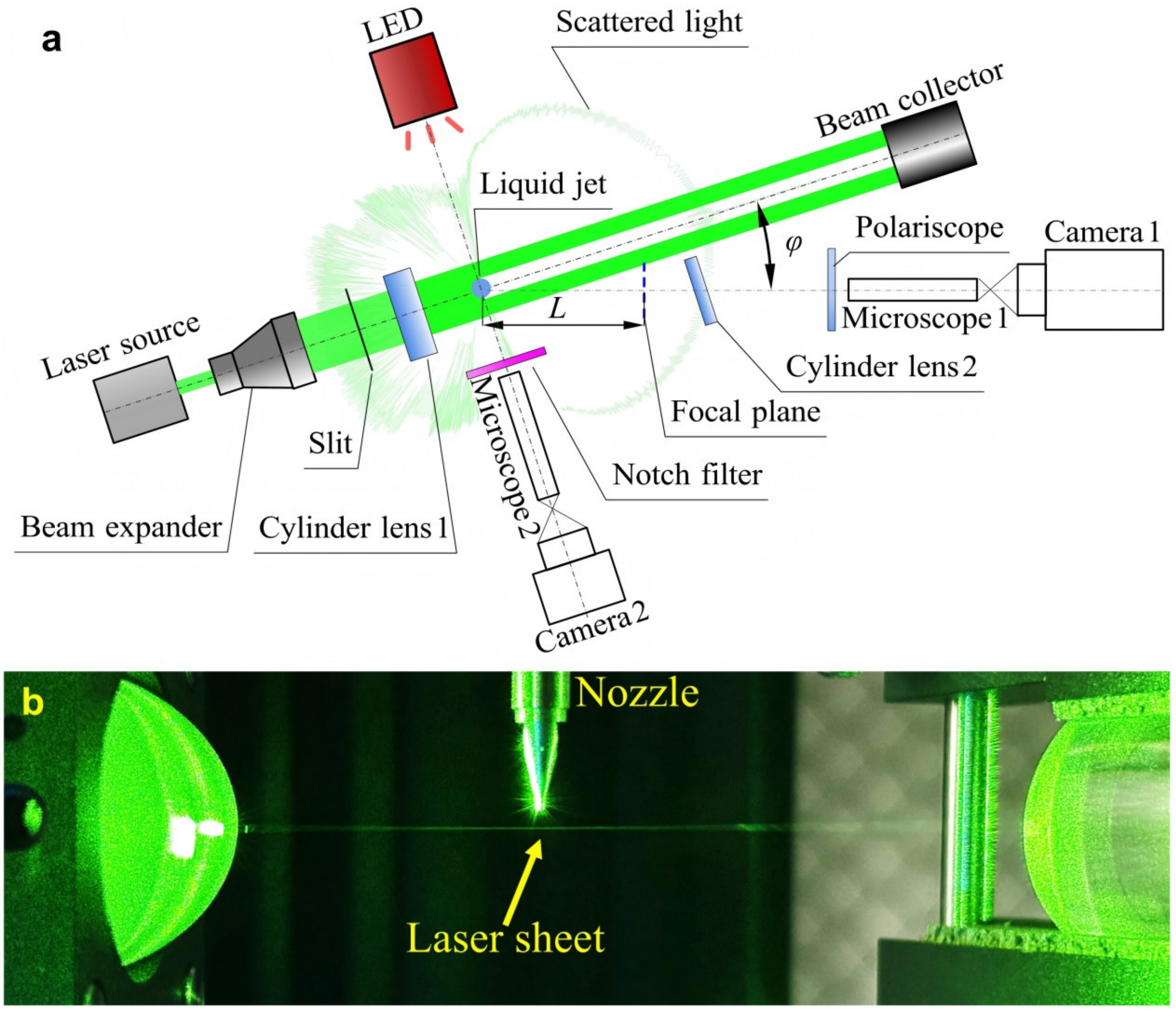

**Fig. 3** Experimental setup for Mie scattering measurements. **a,** Schematic of the optical configuration. **b,** Photograph of the laser sheet.

Fig. 4a displays the captured rainbow scattering fringe pattern. The grayscale values along the marked line are extracted and plotted in Fig. 4b. The light intensity distribution predicted by Lorenz-Mie theory is also plotted for comparison. The experimental signal and the theoretical prediction match closely, demonstrating the feasibility of our experimental setup. After the time-resolved scattering fringe data are collected, they are inverted into jet disturbance signals using an algorithm that will be discussed in Section 2.3.

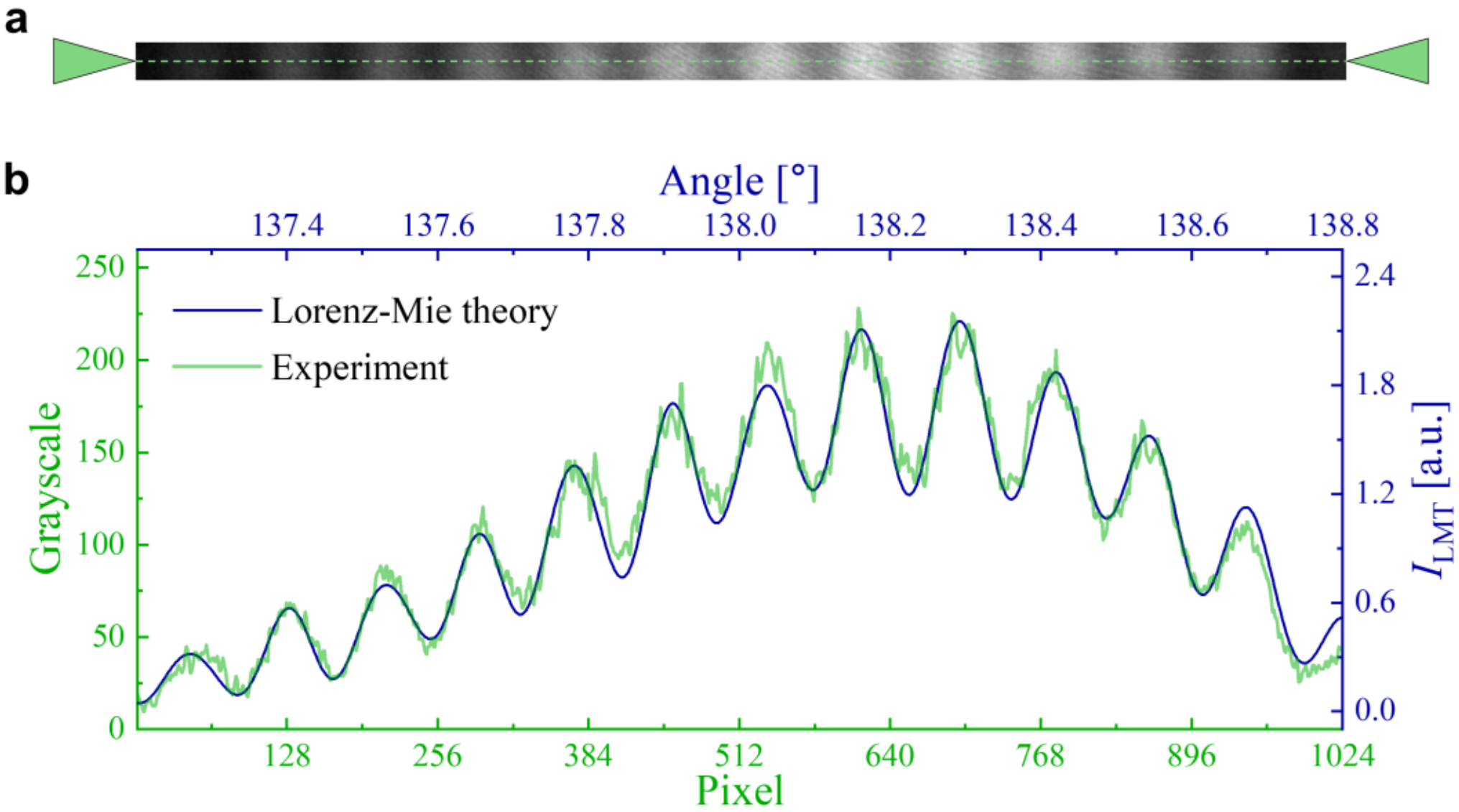

**Fig. 4** Validation of the experimetal setup by comparing measured and theoretical rainbow scattering fringes. **a,** Fringe image. **b,** Extracted grayscale value and theoretically calculated angular intensity distribution.

### 2.2.2 Influence of axial curvature of the liquid jet

In practice, the jet shape may deviate from a uniform cylinder assumed in Lorenz–Mie theory. For example, near the nozzle exits or where surface disturbance amplitudes grow larger at the jet downstream, the jet diameter varies along the axial direction and the axial curvature of the jet diameter also contributes to the scattered light [35]. The extent of this influence depends on the laser sheet thickness. A thinner sheet reduces this effect. In this section, we discuss this phenomenon using an unmodulated laser beam that illuminates the jet as a circular spot 1.5 mm in diameter.

Fig. 5 shows the jet images overlaid with the scattering pattern on the camera screen as the nozzle is elevated to change the laser incident positions ($d_0 \approx 400$ μm). Here the laser is not modulated into a sheet as in Fig. 3. Under this configuration, a scattering fringe with a vertical extent of 1.5 mm would be expected. However, at the first few incident positions, either no scattering fringe is observed or only distorted patterns appear. These phenomena are attributed to jet contraction near the nozzle exit, which causes light rays propagating in different directions to separate [35]. The mechanism of this light separation is discussed in detail in Figs. S4 and S5. From the perspective of spatial disturbance growth, the region closest to the nozzle exit is where the disturbance is the most nascent. Jet contraction, however, forces the expected fringes to appear only at some distance downstream. From another perspective, the fringes produced by Mie scattering can be regarded as the interference between light rays that have traveled along different propagation paths after passing through the jet (see Fig. S4). The axial curvature of the jet gives rise to a series of effects on the light traveling along different paths, including deflection, divergence, convergence, and separation. Although distorted fringes are present in Fig. 5, any particular fringe may well originate from the interference of rays that have propagated from different axial locations along the jet. Consequently, inversion cannot be performed, because the disturbances at those different locations are distinct.

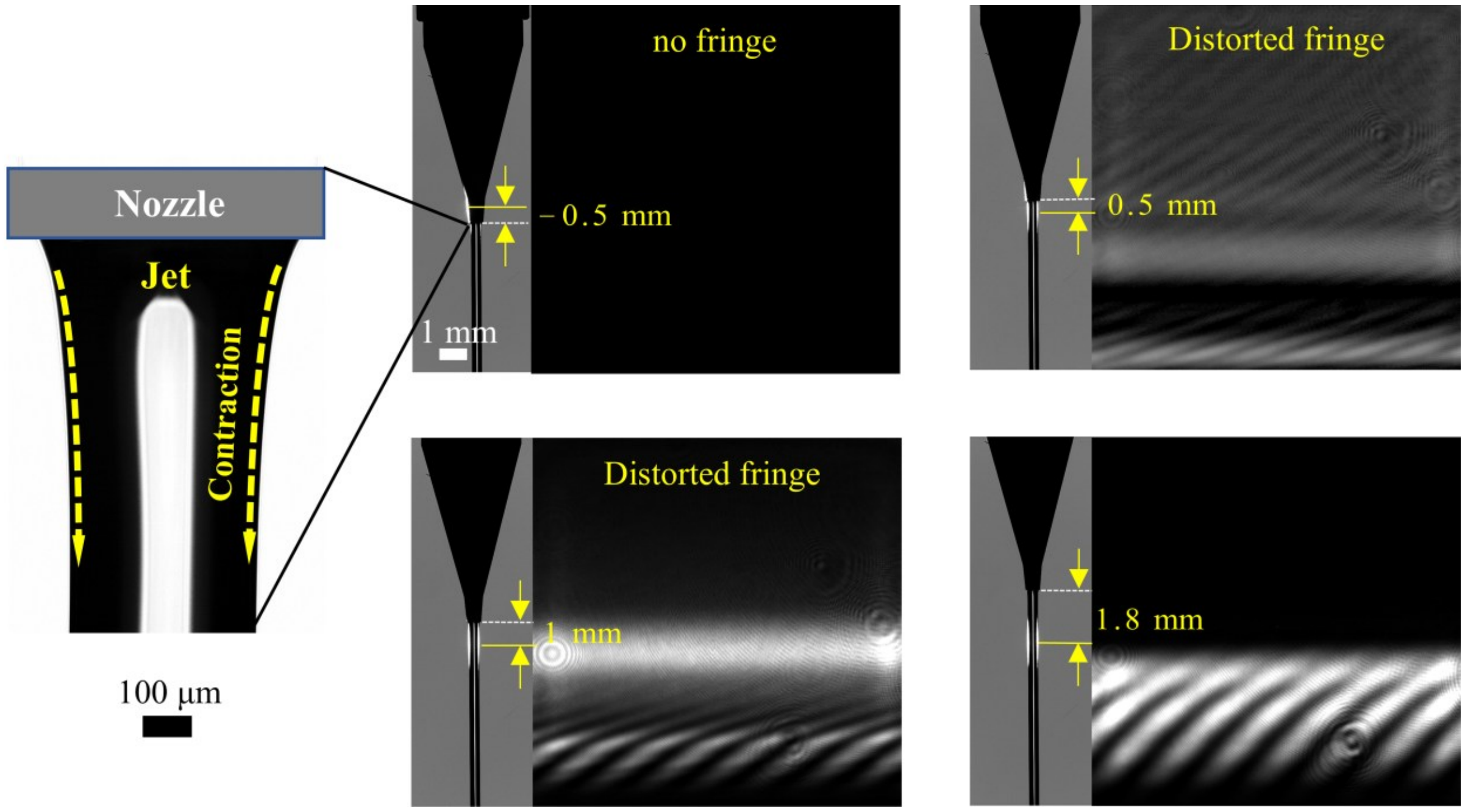

**Fig. 5** The large axial curvature of the jet near the nozzle exit causes the absence or distortion of scattering fringes. Here the unmodulated laser beam of 1.5 mm in diameter is adopted to clearly demonstrate the influence of axial curvature.

In addition to jet contraction at the nozzle exit, large disturbances downstream also distort the scattering fringes under unmodulated laser illumination. Here, "large" refers to an amplitude of above 1 μm, which is shown in an exaggerated scale in Fig. 6 for clarity. As illustrated in Fig. 6, the swollen portion of the jet acts like a convex lens, converging the light rays to form a bright horizontal line on the camera screen; the contracted portion, in contrast, behaves like a concave lens, diverging the light and stretching the fringes axially along the jet image. This lens effect causes the captured scattering pattern to deviate substantially from the prediction of Lorenz–Mie theory, making it impossible to invert scattering variations into jet disturbances. The reason for this impossibility is that the bright horizontal line at the wave crest loses the characteristic alternating dark-bright fringe pattern and can no longer be compared with the Lorenz-Mie prediction. Besides, if the fringes outside the bright line are used for inversion, the retrieved disturbance wavelength becomes distorted as if it has passed through a lens, stretched near the trough and compressed near the crest, thus corrupting the signal.

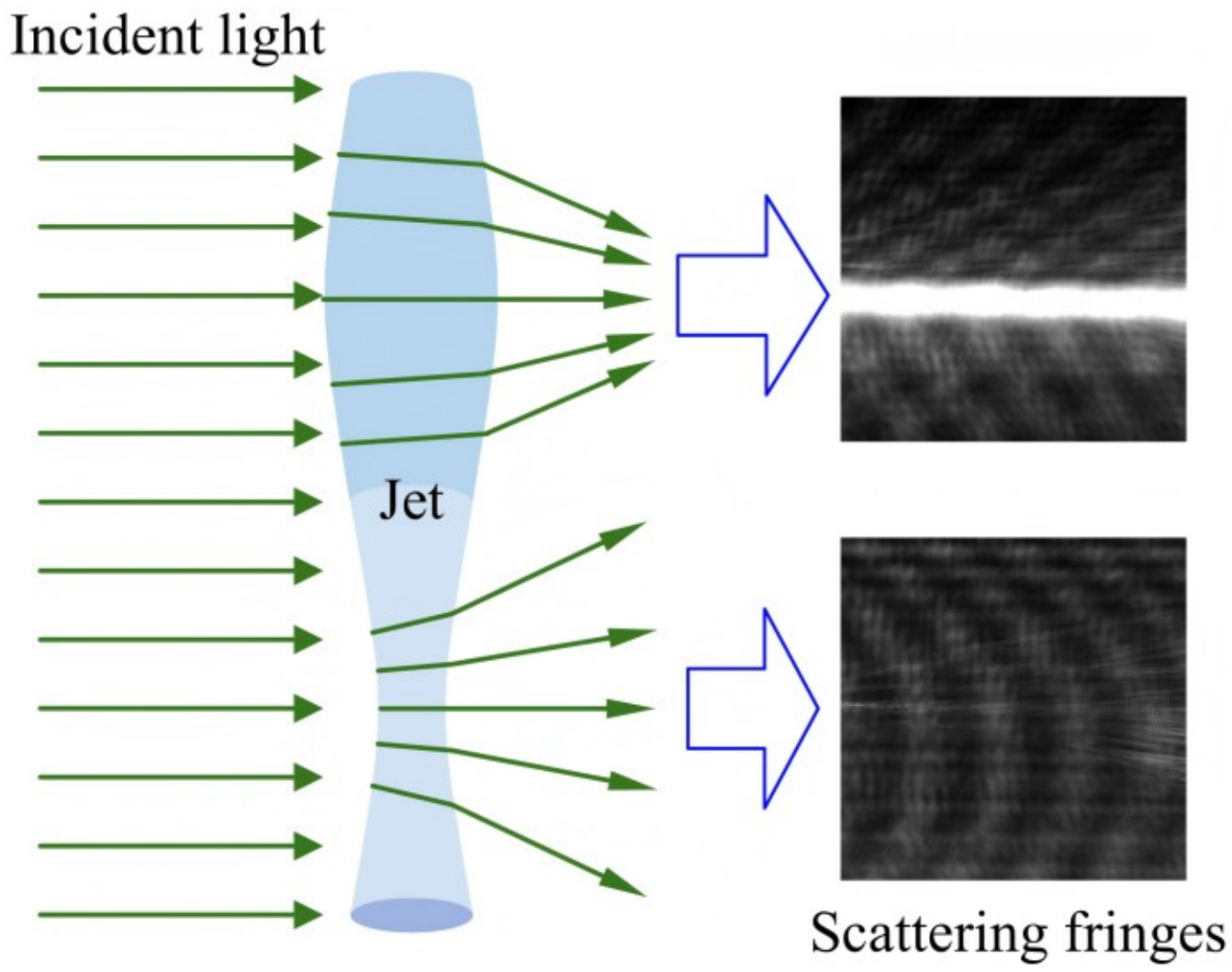

**Fig. 6** Distortion of scattering fringes by the lens effect of large-amplitude jet disturbances under unmodulated illumination.

To address the two aforementioned problems, we modulated the laser into a thin sheet with a thickness of 28 μm (Fig. 3). The optical path illustrated in Fig. 7 shows how the laser sheet propagates and reunites the separated scattered light near the nozzle exit. The forward and rainbow scattering fringes captured under this configuration at $z = L_{\mathrm{j}}/d_0 = 2.5$ (with $d_0 \approx 400$ μm, where $L_{\mathrm{j}}$ is the distance from the nozzle exit) both exhibit the characteristic dark-bright distribution (Fig. 7b, c). Therefore, we successfully achieves the closest feasible measurement position at $z = 2.5$. Since the jet breakup length is large ($L_{\mathrm{b}}/d_0 \gg 2.5$) and disturbances with wavelengths smaller than the jet circumference ($\lambda/d_0 < \pi$) cannot grow, the measurement signals at $z = 2.5$ are sufficient to represent the initial disturbances. Downstream of the jet, the thin laser sheet causes only slight vertical oscillations of the scattering fringes while preserving the theoretically predicted horizontal dark-bright distribution (see Supplementary Video 1). This optical setup provides a maximum measurable amplitude of $\pm 5$ μm in the downstream region. Disturbances larger than $\pm 5$ μm can be accurately measured using the imaging method.

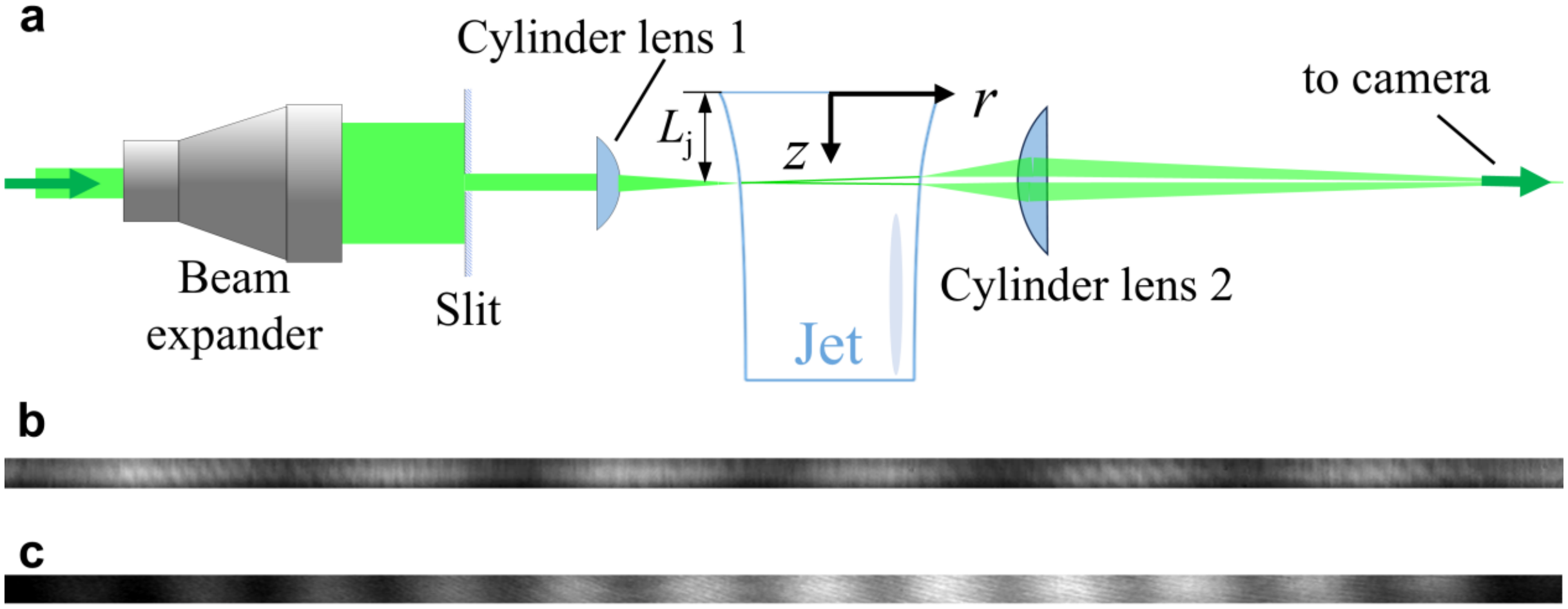

**Fig. 7 a,** Optical path of the modulated thin laser sheet, showing how it reunites the separated scattered light rays. **b,** Forward scattering fringe and **c,** rainbow scattering fringe, both captured at $z = 2.5$ under thin-sheet laser illumination.

### 2.2.3 Influence of jet axis oscillation

Unexpected mechanical vibrations from the platform can trigger oscillation of the jet axis. The displacement amplitude of this axis oscillation may reach the same order as the disturbance amplitude to be measured (~ nm). Such axis oscillation could be amplified by the microscopy imaging system and become mixed with the fringe shifts induced by genuine changes in jet diameter, potentially compromising the measurement. This section evaluates the influence of jet axis oscillation on the measurement of jet surface disturbance

As shown in Fig. 8, when the collection angle is sufficiently small, a lateral shift of the jet axis simply translates the captured fringes without altering their spacing. The relationship between the two displacements is $\delta_2 = \mathrm{M}\delta_1$, where M is the magnification of the microscope, and $\delta_1$, $\delta_2$ are the displacements of the jet axis and the scattering fringes, respectively. In our setup, the maximum M is 18.4 (Leica Z16 APO with 2 × objective lens). By contrast, the equivalent magnification associated with fringe displacement due to jet diameter variation is on the order of $10^3$ to $10^4$ (see Fig. S2), which is overwhelmingly larger than M. Consequently, with proper control of platform vibration [26], the jet axis oscillation introduces negligible noise relative to the signal of interest.

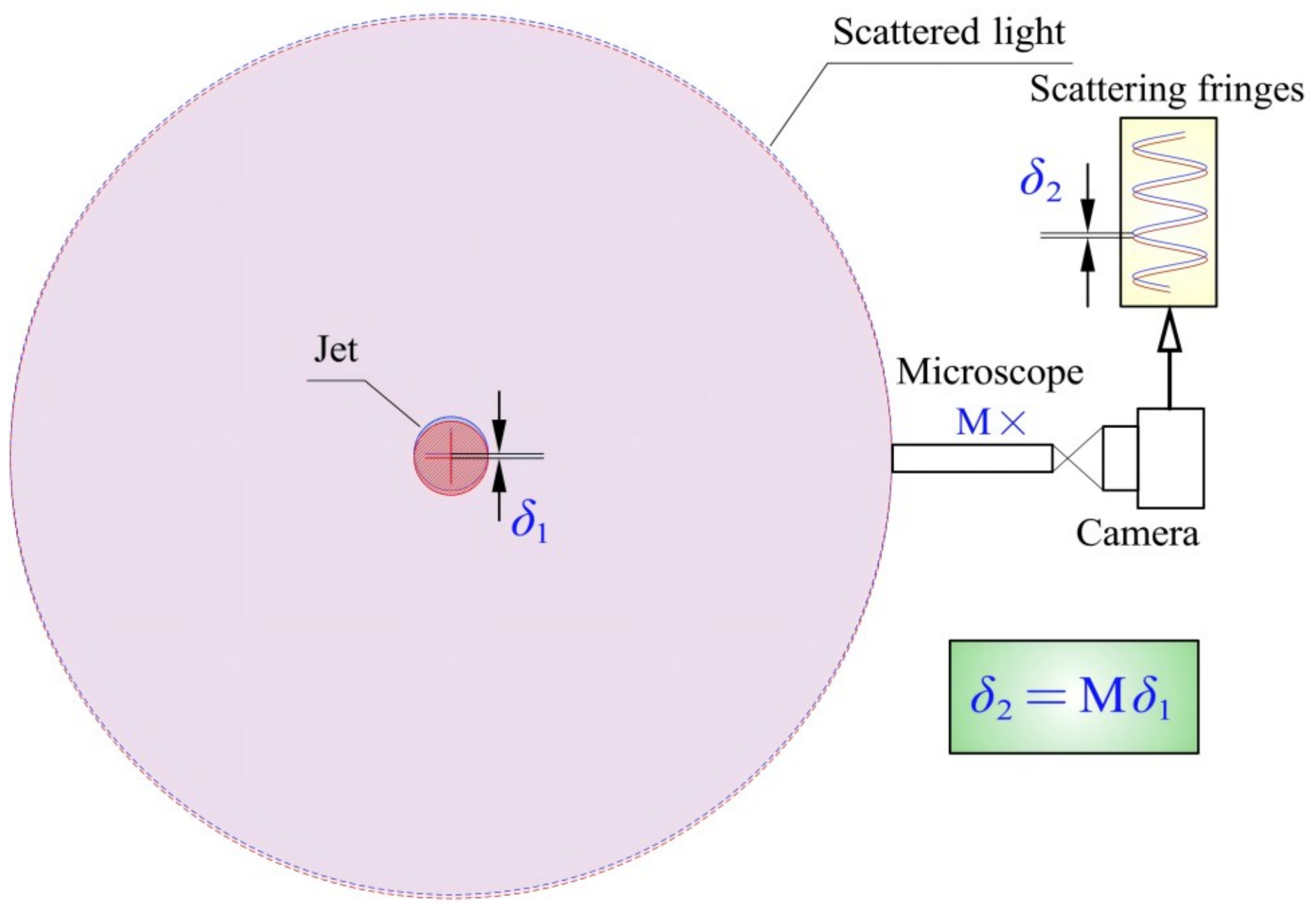

**Fig. 8** Influence of jet axis oscillation on the scattering measurement.

### 2.3 Disturbance inversion algorithm

#### 2.3.1 Basic framework

We propose an algorithm for fringe-to-diameter inversion based on cosine similarity and cross-correlation trajectory, as illustrated in Fig. 9. In step 1, a theoretical database $\mathbf{I}_{\text{thoery}}$ is established by calculating the scattered light intensity distributions of cylinders with different diameters around a reference value $D_0$. The diameter step size is $\alpha$, and the database covers a range from $D_0 - \text{H}\alpha$ to $D_0 + \text{H}\alpha$, where H is a preset constant. $\alpha$ sets the intrinsic resolution of the algorithm, which can be made arbitrarily small, but the true measurement resolution is ultimately bounded by experimental factors. $D_0$ equals to the mean jet diameter $d_0$, which can be obtained from the scattering fringe pattern (see Supplementary Information and Figs. S6, S7). Meanwhile, a set of experimental scattering fringe images is used to produce *n* grayscale vectors corresponding to *n* time instants of the jet disturbance.

Cosine similarity quantifies the similarity between two vectors. In step 2, one grayscale vector from the experimental fringes is selected, and its cosine similarity with each vector in the database is computed, yielding a cosine vector $\mathbf{Cos}(t)$. By repeating this procedure for all *n* grayscale vectors and arranging the resulting $\mathbf{Cos}$ vectors chronologically side by side, a heat map is obtained. Deeper colors in the heat map represent higher similarity, indicating that the experimental data are more consistent with the theoretical calculations for a specific diameter. Consequently, the trajectories in the **Cos** heat map already reveal the variation of the jet diameter.

Since directly extracting trajectories from the **Cos** heat map is somewhat difficult, we introduce a cross-correlation calculation in step 3. Specifically, the relative displacement between each cosine vector and the vector of the first frame is computed via cross-correlation. The resulting **Cor** heat map clearly highlights the temporal change trajectory in its center. Finally, by applying an extreme-value tracing method in step 4, the jet surface disturbance in the form of diameter variation is retrieved.

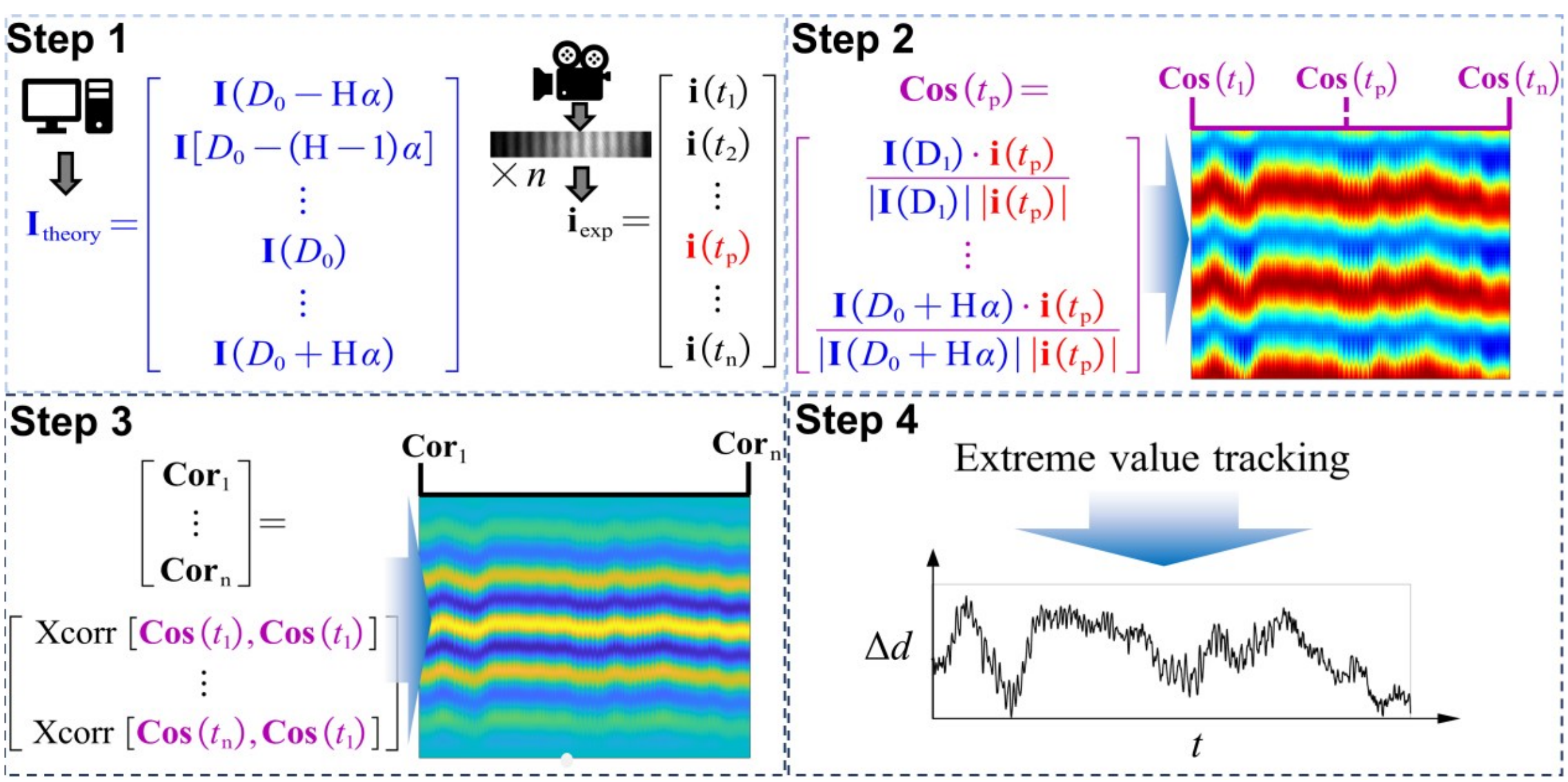

**Fig. 9** Flowchart of the disturbance inversion algorithm

Further algorithm corrections for small (< 10 nm) and larger (> 1 μm) disturbance amplitudes are presented in the next two sections.

### 2.3.2 Algorithm correction for small disturbances

When measuring disturbances very close to the nozzle ($|\Delta d| < 10$ nm), the phase shifts between the **Cos** vectors of adjacent frames become extremely small. In this regime, the method of tracking the offset of extreme points in the **Cor** vectors ceases to be effective. This is illustrated in Fig. 10a, which shows the superimposed **Cor** vectors obtained from 500 consecutive images in a typical experiment. Here, the parameter H is set to 1000, and the Lag index denotes the number of phase-shift steps, with index 2000 corresponding to zero phase shift. Each lag step corresponds to a change in jet diameter of $\alpha$. The magnified view in Fig. 10b reveals that the extremum position remains practically unchanged, and consequently, tracking the peak yields no meaningful disturbance information (Fig. 10c). However, the shoulders of the $\mathbf{Cor}$ curve on either side of the peak exhibit distinct differences (Fig. 10d). Accurate phase shifts can be obtained by locating the intersection of the $\mathbf{Cor}$ curve curve with a horizontal line placed at a level between the maximum and minimum, as indicated by the dashed line in Fig. 10d. By tracking the crossing points at $\mathbf{Cor} = 0.5$, we obtain the reliable disturbance data shown in Fig. 10e. The accuracy of this method is within 7% for very small amplitudes (see the discussion in Supplementary Information for details).

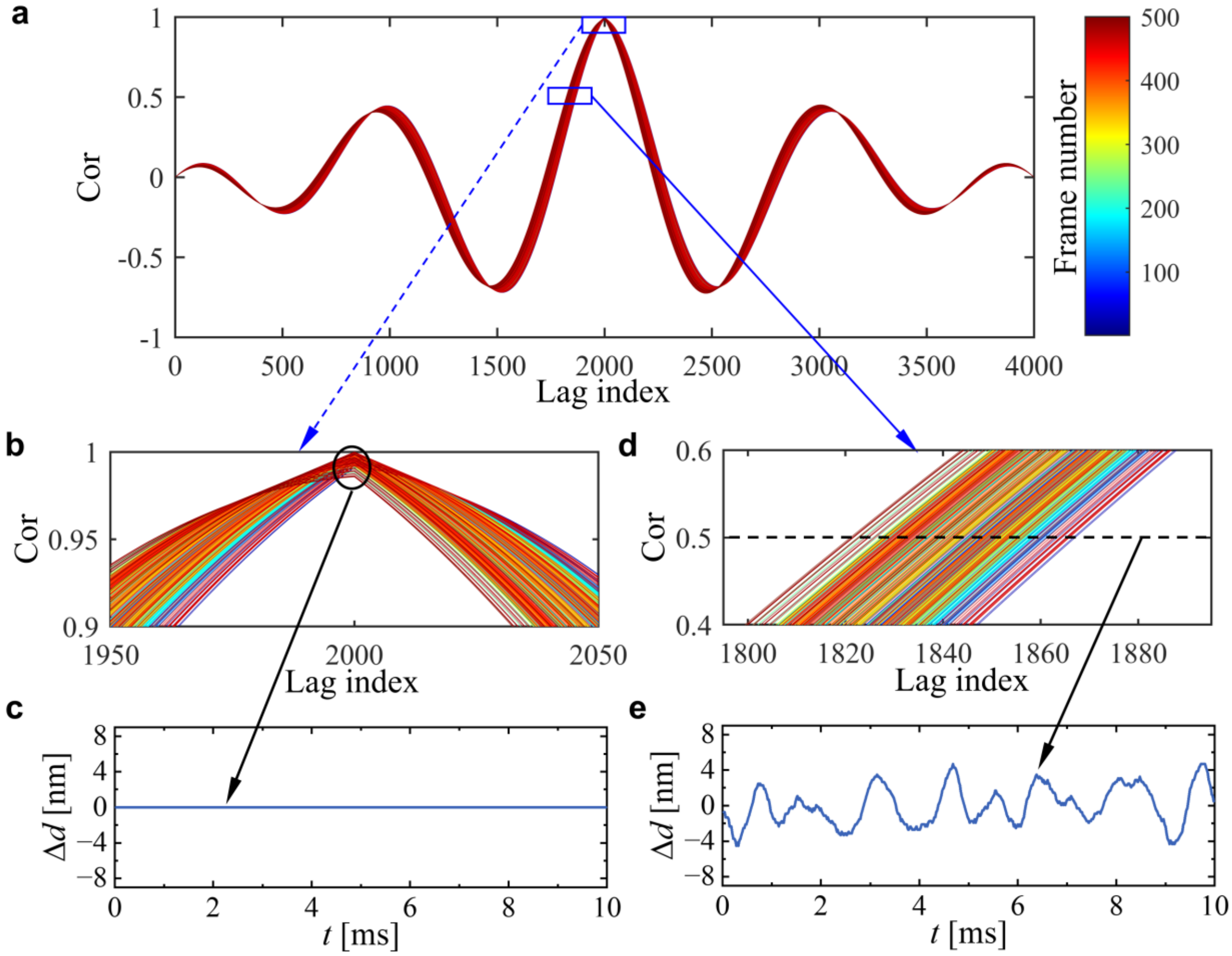

**Fig. 10** Principle and performance of the threshold-crossing method for small disturbances. **a**, Overlay of **Cor** vectors from 500 frames (*Oh*=0.0075, *We*=10, *z*=20). **b**, Magnified region around the extremum. **c**, Result obtained by tracking the peak. **d**, Magnified region of the **Cor** shoulder and the horizontal threshold (dashed line) used for intersection tracking. **e**, Disturbance signal extracted using the threshold-crossing approach, which captures the small variations effectively.

### 2.3.3 Algorithm correction for large disturbances

Localized fringes exhibit periodicity (Fig. 11a) when light is scattered by a large cylinder ($R/\lambda \gg 1$), which introduces difficulties in the inversion of relatively large disturbances ($\Delta d > 1$ μm). As shown in Fig. 11b, when the phase shift between adjacent frames is large, tracking only the $\mathbf{Cor}$ extreme values can cause discontinuities in the trajectory. To correct such discontinuities, we add an additional step after detecting the $\mathbf{Cor}$ extreme value. As illustrated in Fig. 11c, the basic principle is to seek the smoothest path rather than relying solely on the minimum distance between extreme values of adjacent frames. Specifically, for each candidate extreme value in the next frame, the angle between it and the extreme value determined in the previous frame is calculated. The candidate whose angle deviates the least from the angle between the previous two frames is selected.

The corrected trajectory in Fig. 11d exhibits excellent smoothness and faithfully reflects the actual disturbance morphology. This angle-judgment step extends the measurement range to $\pm 5$ μm, which already surpasses the minimum resolvable scale of microscopy imaging. Consequently, by combining the scattering method with imaging, disturbances over nearly the entire jet can be accurately quantified.

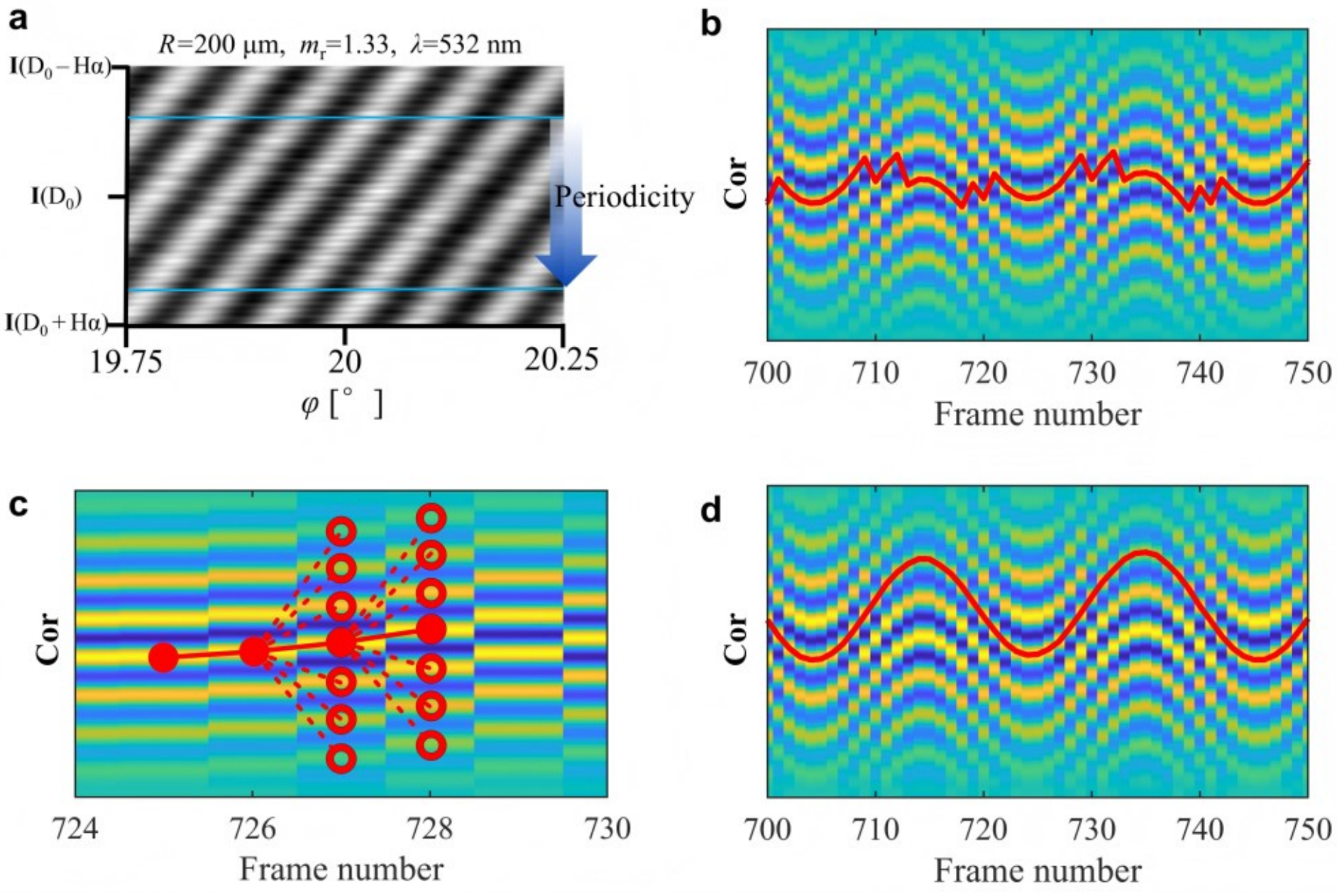

**Fig. 11** Correction of the inversion algorithm under large disturbances. **a**, Periodic localized fringes for a large cylinder. **b**, Discontinuous trajectory from direct extreme-value tracking. **c**, Angle-based selection of the smooest path. **d**, Corrected trajectory.

To comprehensively evaluate the accuracy and practical applicability of the proposed measurement method, we present a detailed analysis in the Supplementary Information and Figs. S8-S12, covering the entire error propagtion chain and a numerical simulation framework based on noise-added theoretical fringes. The simulation results demonstrate that our method achieves high accuracy, with sub-nanometer resolution.

## 3 Noise characterization and suppression

Since the disturbance signals of interest are at the nanometer level, rigorous noise characterization and suppression are essential to prevent misinterpretation of environmental artifacts as physical jet dynamics.

In our early experiments, a motorized translation stage was used for precise nozzle positioning, which introduced significant noises into the initial disturbance measurements at $z$=2.5 (Fig. 12a). The measured disturbance amplitude falls within ±50 nm, and the Fast Fourier Transform (FFT) spectrum exhibits peaks at 260 Hz, 400 Hz, 500 Hz and 770 Hz.

After suppressing the stage-induced vibration, the initial disturbance signal Fig. 12b still exhibited noise-like characteristics with frequency peaks at 50 Hz and 114 Hz. The 50 Hz component originates from ground vibrations, while the 114 Hz peak is generated by the camera cooling fan. To eliminate these contributions, we mounted the entire measurement system on an air-bearing platform and turned off the camera fan during acquisition.

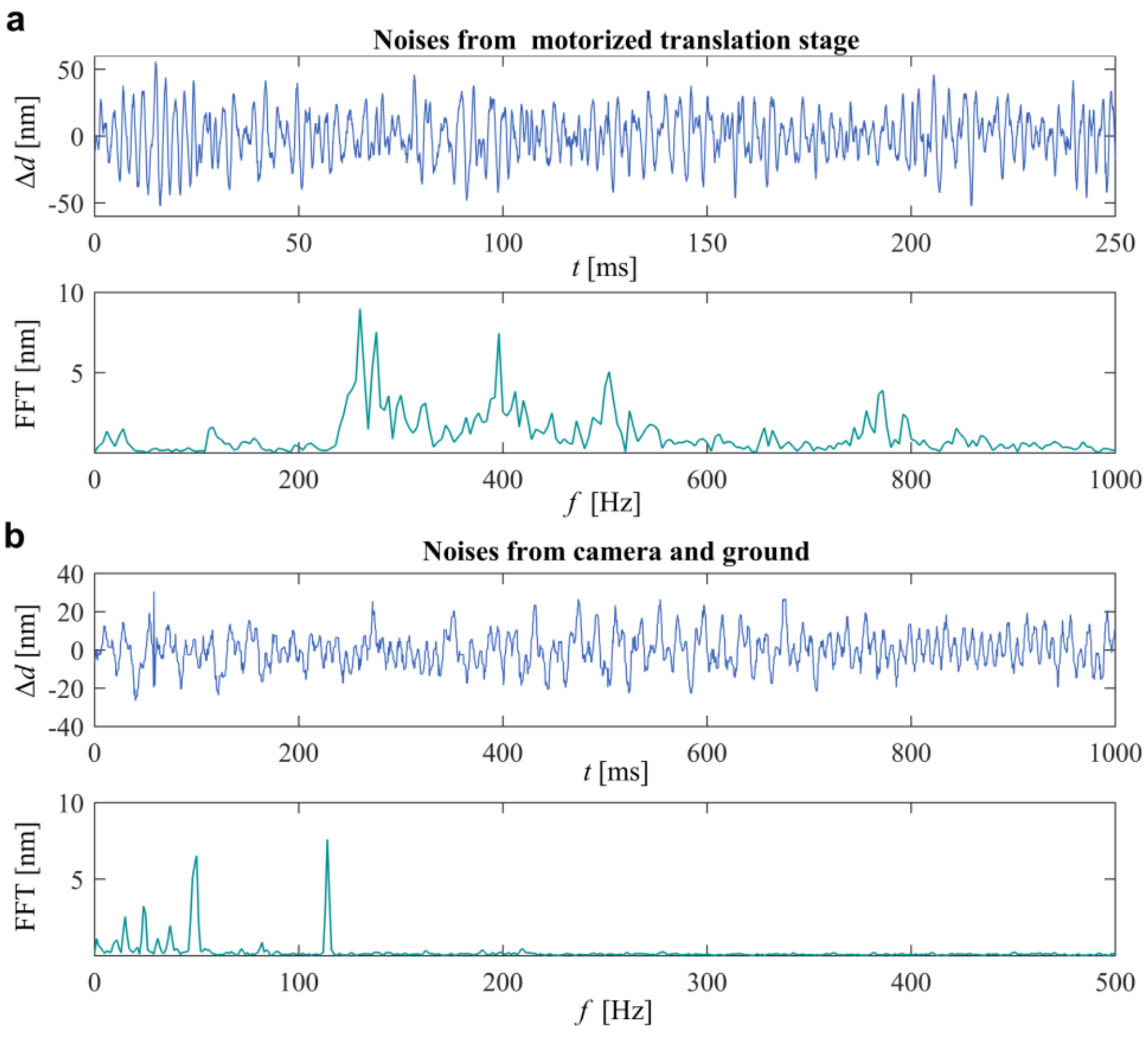

**Fig. 12** Background noises in the initial disturbance measurement at $z$ = 2.5. The noises were identified from **a**, motorized translation stage and **b**, camera and ground. ($Oh$=0.0075, $We$=10).

With the disturbance noise significantly reduced, we next evaluated the ultimate background noise of our system by replacing the nozzle with a glass rod (~380 μm in diameter) to simulate a static jet with no diameter variation (Fig. 13a). Under this condition, the measured signal reflects the combined effects of unknown mechanical vibrations, laser wavelength stability, air flow disturbances, and noises introduced in our inversion algorithm (see Supplementary Information and Fig. S8 for details). By applying the same procedures described in Fig. 9, we extracted the background noise signal from this static configuration in Fig. 13b. Given its random nature, we characterized it by computing the power spectral density (PSD), which is appropriate for stochastic signals. The noise amplitude spans a range of $\pm 0.5$ nm, corresponding to an RMS (root mean square) value of 0.18 nm. The PSD decays gradually with increasing frequency and exhibits no distinct peaks, suggesting white-noise-like behavior and consistent with the PSD obtained from our numerical simulation in Fig. S9. This sub-nanometer noise floor represents the ultimate background noise of our system and highlights the excellent performance of our measurement system as well as the exceptionally high precision of the scattering-based method.

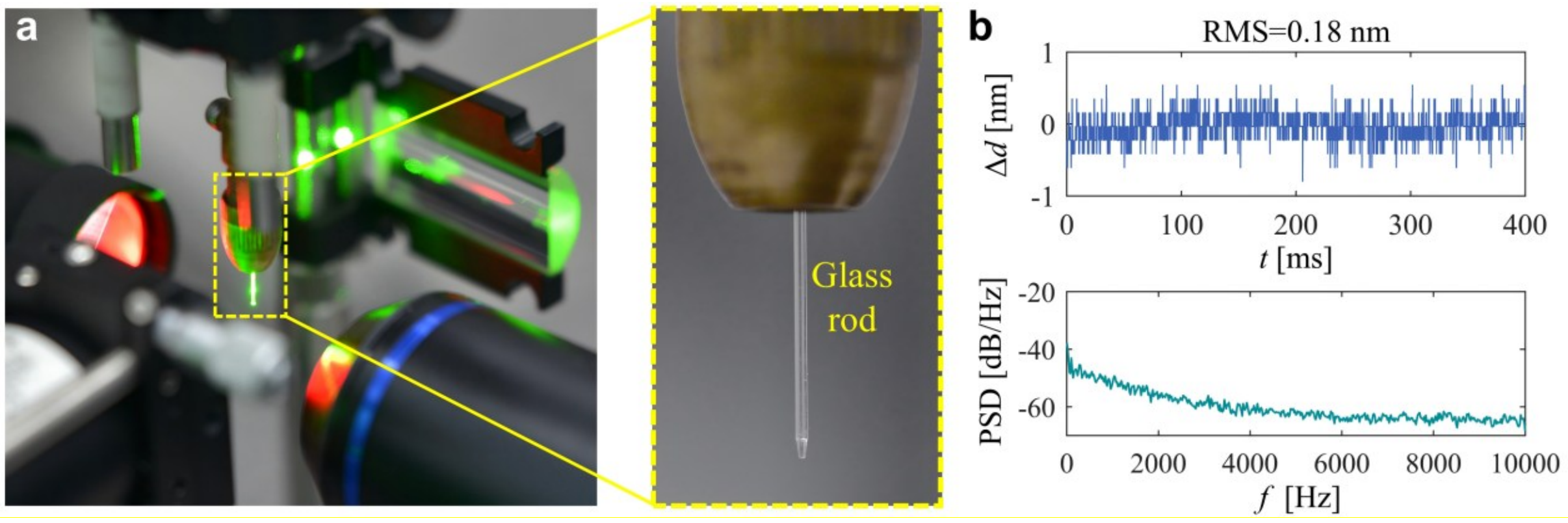

**Fig. 13** Background noise floor assessment of the measurement system. **a,** A static glass rod replaces the jet to eliminate diameter variations. **b,** Extracted background noise signal and its PSD.

## 4 Validation of the measurement method

### 4.1 Axisymmetry verification of the disturbances

The Lorenz-Mie theory assumes axial symmetry of the jet, which may be invalid in experiments as the initial disturbances may distribute at both the circumferential and axial directions. Fortunately, the circumferential disturbance always decays [38], and the jet becomes axisymmetric below a certain axial position. To ensure the validation of the measurement of our Mie scattering method, the asymmetry of the disturbance should be checked before experiments.

Since scattered light is generated over the full 360° around the jet, measurements taken at different azimuthal angles should yield identical disturbance profiles if the disturbances are indeed axisymmetric. To verify this, we compare the signals obtained from forward scattering ($\varphi = 20°$) and rainbow scattering ($\varphi = 138°$). At $z = 2.5$ (Fig. 14a), although the local details of the two time traces differ slightly, their overall trends are in good agreement. The corresponding PSDs also show reasonable consistency up to 5000 Hz, though the PSD level for rainbow scattering is somewhat lower than that for forward scattering. We attribute this minor discrepancy to the difference in the noise floors inherent to the two scattering configurations, given that the disturbance amplitude at this position is extremely small (see the detailed precision analysis in the Supplementary Information).

At $z$ = 7.5 (Fig. 14b) and $z$ = 12.5 (Fig. 14c), the time traces exhibit excellent consistency, particularly in the occasional localized large-amplitude pulses, and the PSDs match closely across the entire frequency range up to 5000 Hz. These results provide strong evidence that our initial disturbances measurements satisfy the axisymmetry assumption, which is fundamental to the analysis presented in the following sections. Moreover, the PSD comparison suggests that jet signals with spectral magnitudes above approximately −50 dB/Hz can be reliably regarded as axisymmetric within the present experimental configuration.

Overall, we consider that from $z$ = 2.5 onward, the jet already exhibits essentially axisymmetric behavior. The measured Mie scattering data are thus physically meaningful and can accurately capture the frequency-domain information of the unstable modes of interest. Consequently, in the subsequent analysis, we use the data acquired at $z$ = 2.5 as the initial disturbance for further investigations.

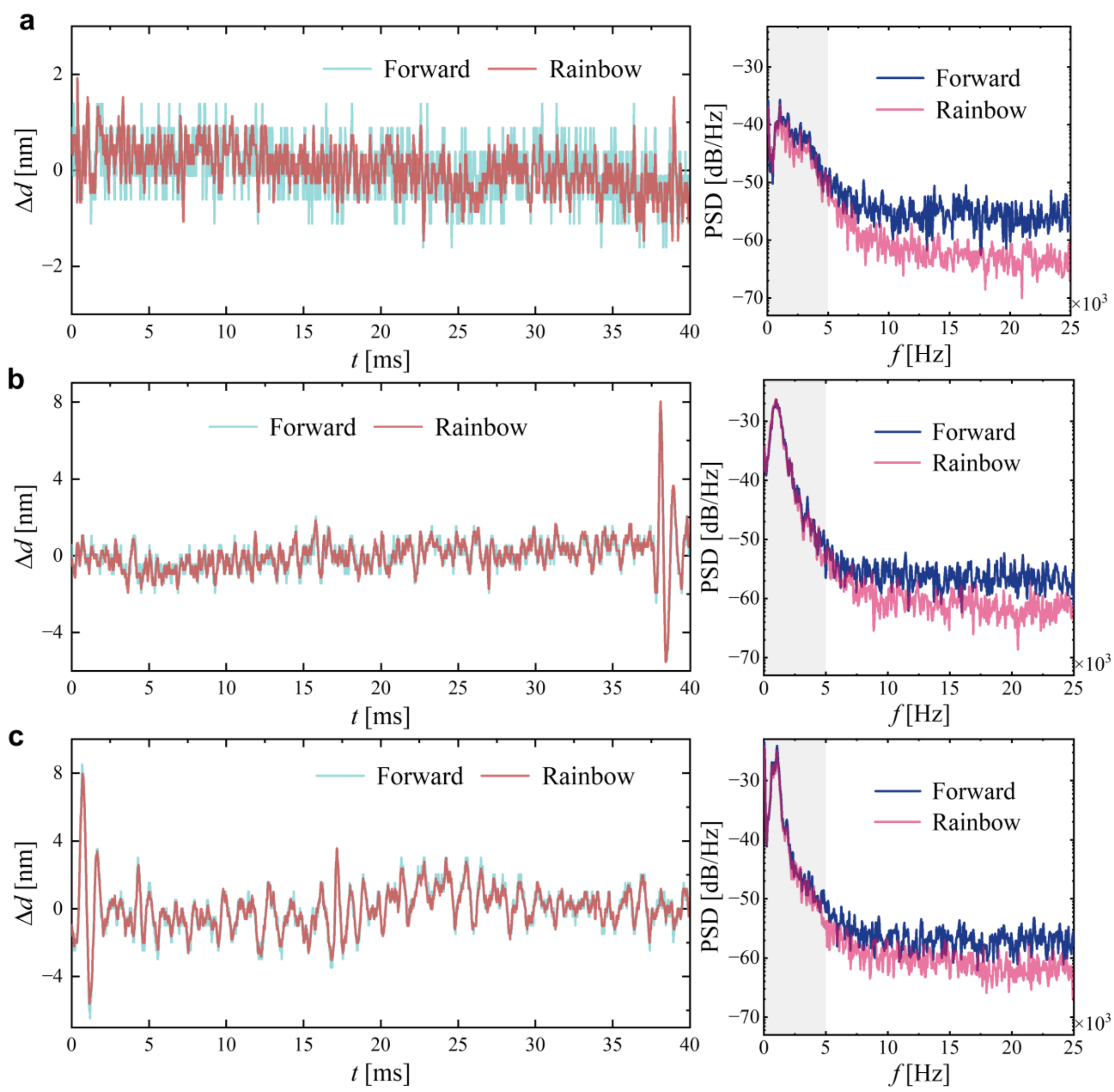

**Fig. 14** Verification of axial symmetry of jet disturbances. Disturbances profiles measured simultaneously by for forward scattering and rainbow scattering at three axial positions: **a**, z=2.5, **b**, z=7.5, **c**, z=12.5. The shaded region in the PSD shows good consistency. (*Oh*=0.0075, *We*=10).

## 4.2 Comparison with synchronous imaging

To validate that the signals obtained from Mie scattering indeed represent jet disturbances, a single axial position on a jet was selected and measured using both the scattering and the imaging method synchronously. This position was chosen such that the disturbance amplitude fell within the measurement range of both methods, which is approximately $\pm 2$ μm. No external excitation was applied to the water jet, ensuring that the validation signal was entirely random. The imaging signal was acquired with a 20 × microscope, corresponding to a pixel-level resolution of 1 μm. By further applying a sub-pixel edge extraction algorithm [39], the effective resolution was improved, as illustrated in Fig. 15. The disturbances measured by Mie scattering and by imaging show good agreement in overall trend, with the scattering signal exhibiting a smoother profile and lower apparent noise. Most of the deviations between the two data sets lie within 0.1%. These discrepancies are primarily attributed to the imaging method, including the raw image quality and the inherent limitations of the sub-pixel algorithm.

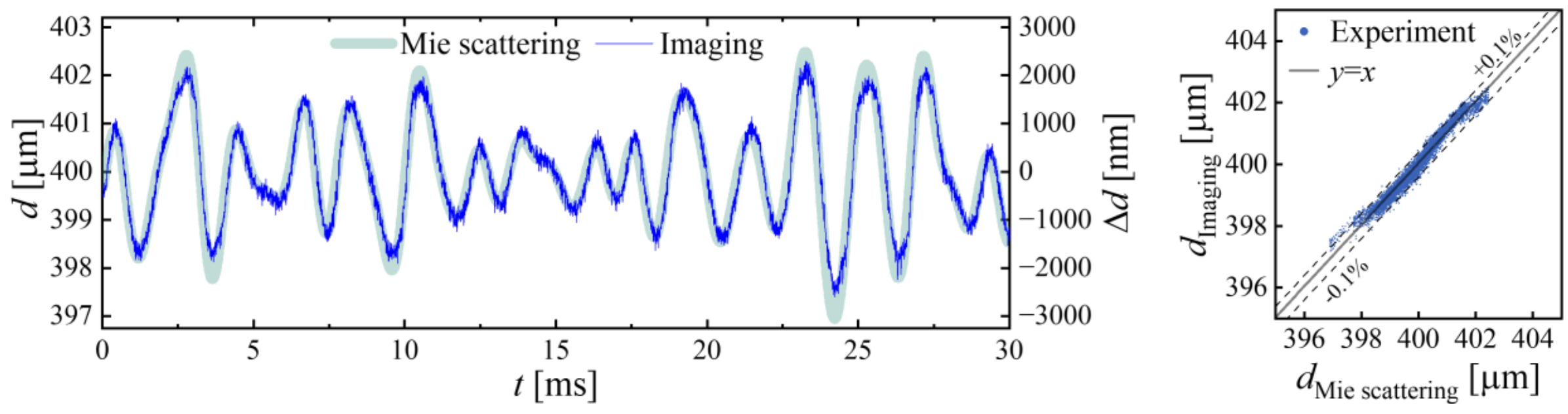

**Fig. 15** Validation of the scattering method against synchronous imaging measurements (*We*=4.3, *Oh*=0.0075, *z*=20).

### 4.3 Comparison with scanning electron microscopy

Inspired by the work of Warke and Giessen [40], we compared the Mie scattering method with scanning electron microscopy (SEM) using a quartz fiber sample approximately 9 μm in diameter and 6.6 mm in length (Fig. 16a). SEM images were acquired with a Thermo Fisher Apreo 2 at a magnification of 40,000 ×. The sample was made in-house from quartz wool (see Fig. S13), and the diameter variation along the fiber was measured by both Mie scattering and SEM. Fig. 16b presents the comparison for one representative sample, selected from a set of measurements for its relatively regular diameter variation. The two methods show good agreement in overall trend at the nanoscale level. The deviations can be attributed primarily to surface impurities (see Fig. S14 for SEM images of the two points with the largest error margins), as the scattering signal is insensitive to such impurities. In addition, possible non-axisymmetry of the fiber cross-section may also contribute to the deviation, since the scattering method inherently assumes a perfectly uniform cylinder.

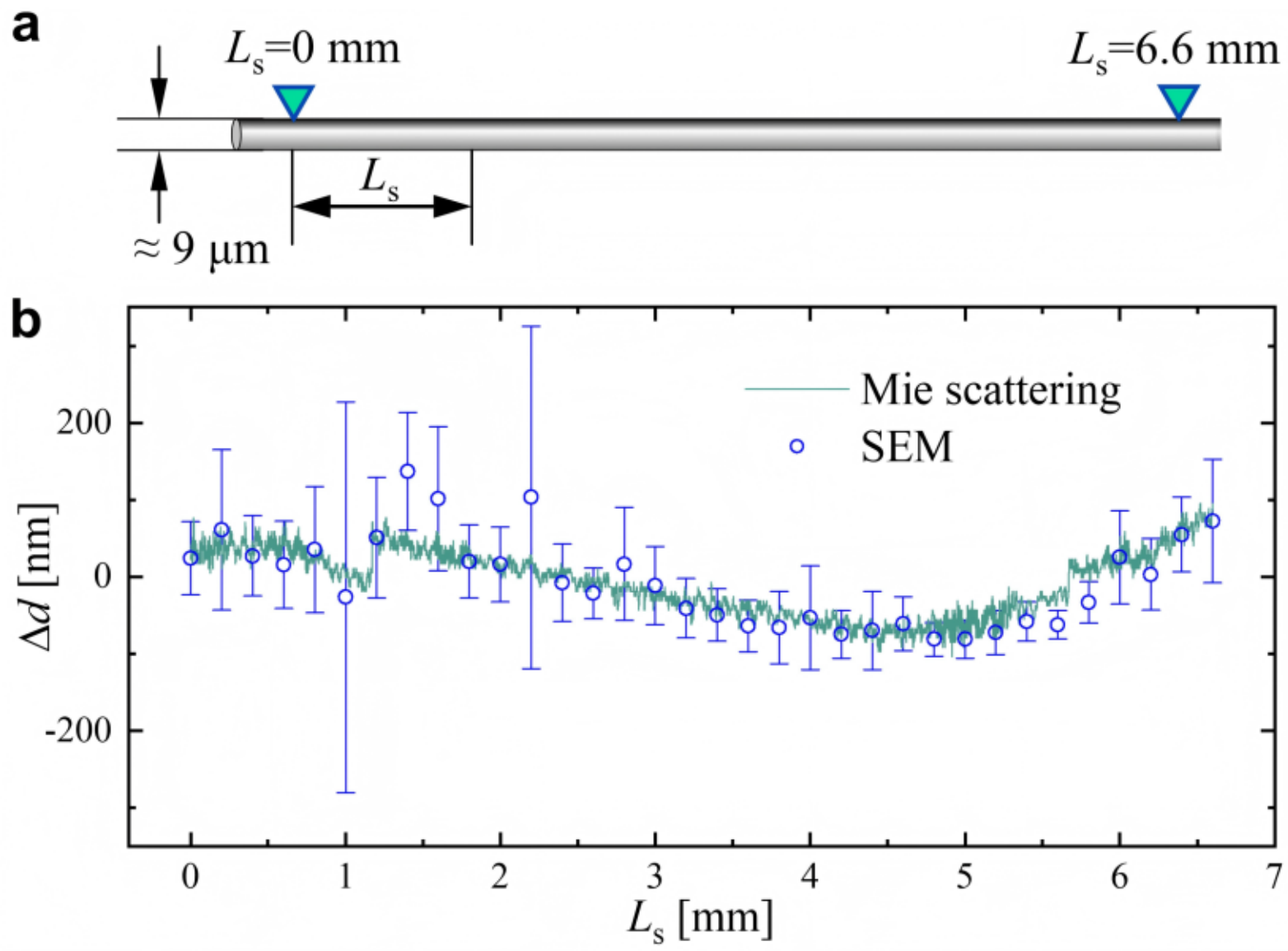

**Fig. 16** Comparison of diameter variation measurements between Mie scattering and SEM. **a,** Schematic of the quartz fiber sample used for the comparison. **b,** Diameter profiles along the fiber measured by both methods.

These two independent validations, including time-resolved imaging and static SEM, collectively confirm that the Mie scattering method reliably captures the dynamic nanoscale jet disturbances. We thus consider the method sufficiently accurate for the subsequent measurements in this study.

## 5 Application of the measurement method

With the measurement accuracy and robustness established in the preceding sections, we now apply the method to characterize the initial disturbances for water jets in the near-nozzle region. Fig. 17 presents examples of initial disturbance measurements at $z$=2.5 for water jets at $Oh$=0.0075 and various Weber numbers. The PSD of the background noise is also shown for comparison, confirming a good signal-to-noise ratio for the measured initial disturbance data. Owing to the inherently random nature of the initial disturbances, the results in Fig. 17 do not cover all possible cases. Nevertheless, two particular characteristics were consistently observed in all our experiments. The first is the occurrence of random pulses (see Fig. 17b-c). These pulses give rise to flat bulges in the PSDs, as highlighted by arrows. The central frequencies of these bulges increase with the Weber number and differ from the dominant frequencies predicted by linear stability analysis, which are marked by dashed lines as $f_d$ (the calculation method for $f_d$ is provided in Fig. S15). The Second characteristic is the presence of low-frequency oscillations, which are more evident in Fig. 17a, b and d. The low frequency oscillation pattern varies across experimental runs, making it difficult to summarize and derive a general law. However, through further investigation we have found some evidence related to these two characteristics.

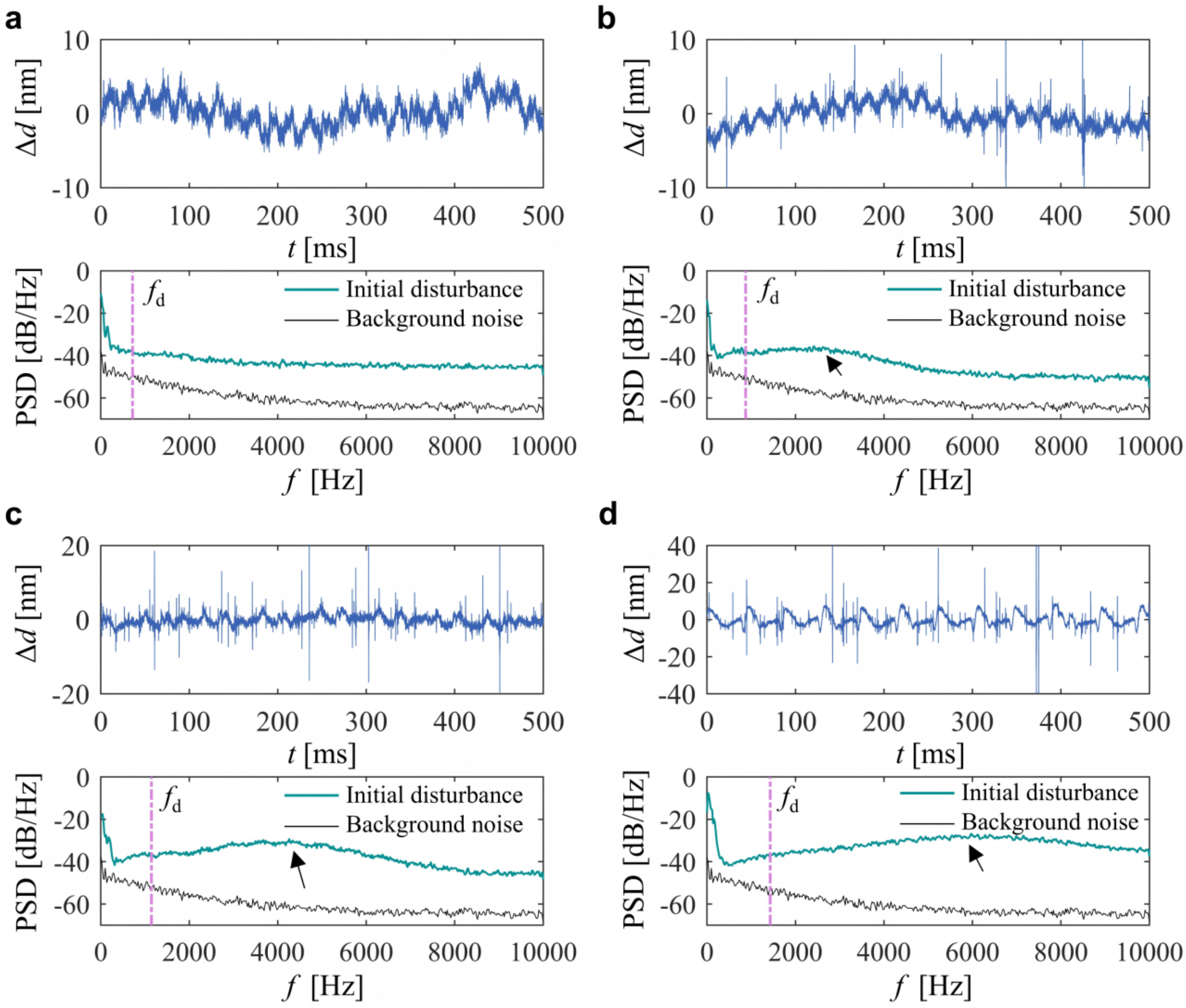

**Fig. 17** Initial disturbance measurements at $z$=2.5 for water jets with $Oh$=0.0075 and various Weber numbers. **a,** $We$=3.2. **b,** $We$=5.9. **c,** $We$=10.9. **d,** $We$=17.3.

Pulse signals exert a strong influence on jet instability. To investigate this effect, we performed synchronous measurements at $z$=2.5 (Mie scattering) and $z$=42.5 (imaging). As seen in Fig. 18, the pulses captured at $z$=2.5 (Fig. 18a) develop into distinct wave packets at $z$=42.5 (Fig. 18b). From a frequency-domain perspective, the broadband signal observed near the nozzle is subsequently shaped by the jet instability dynamics, resulting in a PSD at $z$=42.5 that clearly exhibits the dominant frequency and bandwidth characteristics predicted by linear stability analysis. The pulses thus function as seeds that can trigger the growth of multiple instability wavelengths. This behavior aligns perfectly with the fundamental signal-processing principle that a linear time-invariant system can be fully described by its impulse response. In contrast, the low-frequency oscillations exert little influence on jet stability, owing to the low growth rates at low frequencies.

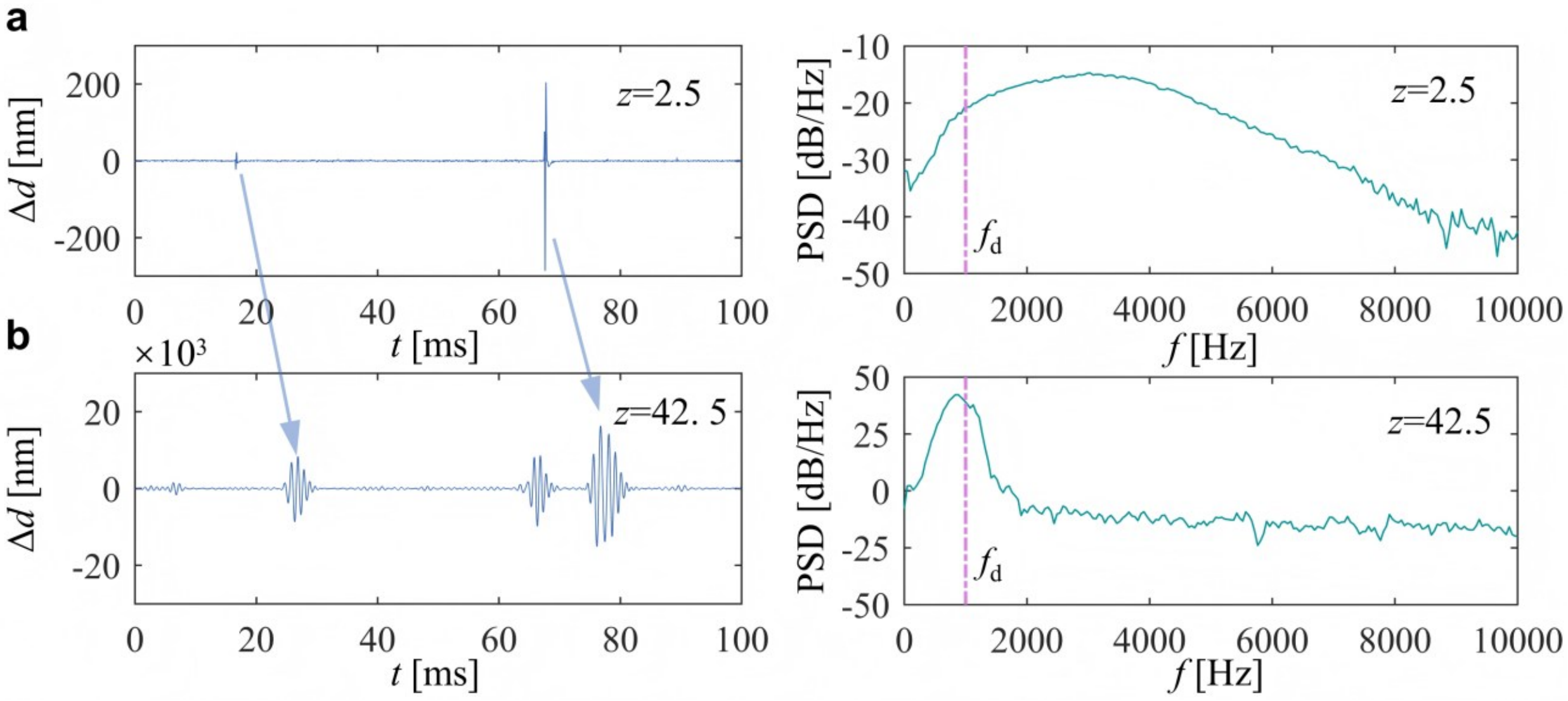

**Fig. 18** Propagation and evolution of pulse signals along the jet. **a,** Pulses measured by Mie scattering at z=2.5. **b,** Distinct wave packets developed at z=42.5 (*Oh*=0.0075, *We*= 9.8).

The underlying mechanisms governing the pulses and the low-frequency oscillations require further investigation. Nevertheless, some evidence has been identified in Figs. S16 and S17. Regarding the pulse signals, nanobubbles formed during injection appear to be the primary cause (Fig. S16). Even after careful filtration [26] to remove bubbles before the experiments, the pulses could not be completely eliminated, suggesting that they are likely an intrinsic feature of water jets. In addition to nanobubbles, a meniscus exists at the nozzle exit, anchored to the three-phase contact line. Flow instabilities within this meniscus may introduce low-frequency oscillations into the jet, as initially examined through a comparison in Fig. S17. Overall, these newly observed phenomena call for further dedicated studies to be fully clarified.

## 6 Conclusions

We have developed a Mie scattering-based measurement system for tracking nanoscale diameter variations in liquid jets, comprising a dedicated optical setup and a robust fringe-to-diameter inversion algorithm. The system's accuracy is ensured by several key design features: thin-laser-sheet illumination that effectively suppresses distortions from jet contraction and the lens effect of large-amplitude disturbances; a negligible influence from jet axis oscillation, as the equivalent magnification for diameter-induced fringe shifts far exceeds the optical magnification; and a remarkably low background noise floor of 0.18 nm RMS with white-noise characteristics, as assessed using a static glass rod. Symmetry tests demonstrate that the disturbances are axially symmetric at near-nozzle regions, validating the underlying theoretical assumptions of Mie scattering. Cross-validation with synchronous imaging and scanning electron microscopy confirms the method's quantitative reliability.

Using this technique, we have characterized the initial disturbances at a position of 2.5 times jet diameter below the exit for water jets. Two distinct stochastic features are consistently observed: (a) broadband pulses, which are shown to serve as seeds that trigger a spectrum of unstable frequencies and evolve into wave packets downstream; and (b) low-frequency oscillations, which contribute little to the eventual breakup because of the low growth rates at those frequencies.

Our findings not only fill the long-standing gap in measuring initial nanoscale disturbances but also provide primary experimental measurement for the early onset of water jet instability. Besides, our method can be readily extended to other cylindrical micro-objects for high-precision diameter fluctuation measurements. This work establishes a critical experimental foundation for future studies on jet breakup, receptivity, and active control.

## Acknowledgements

We acknowledge support from the National Natural Science Foundation of China (Grant No. 12502287, U2341281, 12272026) and the Fundamental Research Funds for the Central Universities.

## Author contributions

L. Y. and B. J. conceived the research. L. Y. and B. J. supervised the research. D. Z. carried out the experiments, collected data and analyzed the experimental results with the help of Q. L. and H. S. D. Z., B. J. and L. Y. draft the manuscript. All authors contributed to the interpretation and drafting of the paper.

## Competing interests

The authors declare no competing interests.

# Measurement of Nanoscale Surface Disturbances in Liquid Jets Using Mie Scattering Method

## ——Supplementary information

Dingwei Zhang[1], Qiyou Liu[1], Hu Sun[1], Qingfei Fu[1,2,3], Bingqiang Ji[1,2,4*], Lijun Yang[1*]

[1] School of Astronautics, Beihang University, Beijing 100191, PR China

[2] Aircraft and Propulsion Laboratory, Ningbo Institute of Technology, Beihang University, Ningbo 315800, PR China

[3] National Key Laboratory of Aerospace Liquid Propulsion, Xi'an, 710100, PR China

[4] State Key Laboratory of High-Efficiency Reusable Aerospace Transportation Technology, Beijing 100191, PR China

**Corresponding authors:**

*bingqiangji@buaa.edu.cn (B.J.); *yanglijun@buaa.edu.cn (L.Y.)

## Resolution and equivalent magnification estimation

The resolution is estimated as illustrated in Fig. S1. First, a set of light intensity distribution based on Lorenz-Mie theory are calculated for radii ranging from 400 to 400.4 μm and arranged vertically. The refractive index of water is used. We tracked the movement of one intensity peak as the jet diameter increased by 0.4 μm, and the angular separation between the peak positions is approximately 0.23°. Assuming a camera with 1000 pixels is used to capture the scattered light over the angular range from 137.5° to 138.5°, and since the range is sufficiently small, the arc length is essentially equal to the chord length, the angular resolution is 0.001°. Given that the peak shift is proportional to the radius change, the radius measurement resolution can be deduced as:

$$\delta D = \frac{\Delta D}{\Delta \varphi}\delta\varphi = \frac{400}{0.23}\times 0.001 = 1.7\ \text{nm} \tag{S1}$$

This nanometer-scale resolution estimate dramatically strengthens our confidence in resolving nanoscale disturbances within liquid jets. Additionally, the resolution can be further enhanced by adopting a more sophisticated phase-shift detection algorithm, as shown in Fig. 9 in the main text.

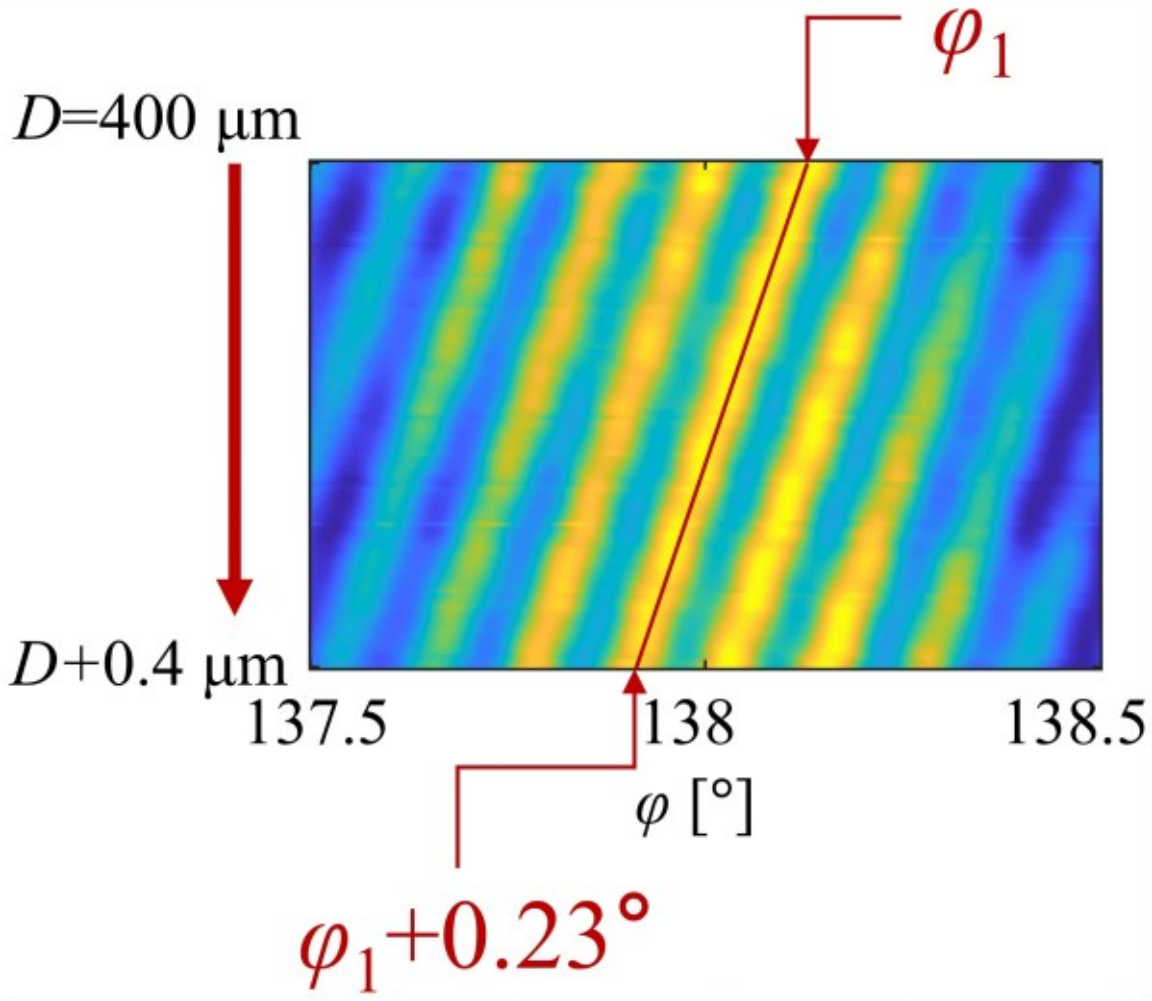

**Fig. S1** An estimation of the measurement resolution

In optical microscopy, the magnification of an objective lens is defined as the ratio of the image size to the object size. By analogy, we define an equivalent magnification for our scattering-based measurement, namely the ratio of the fringe shift on the camera sensor to the actual variation in jet diameter that causes it. An estimation of the equivalent magnifications of both forward scattering and rainbow scattering are shown in Fig. S2.

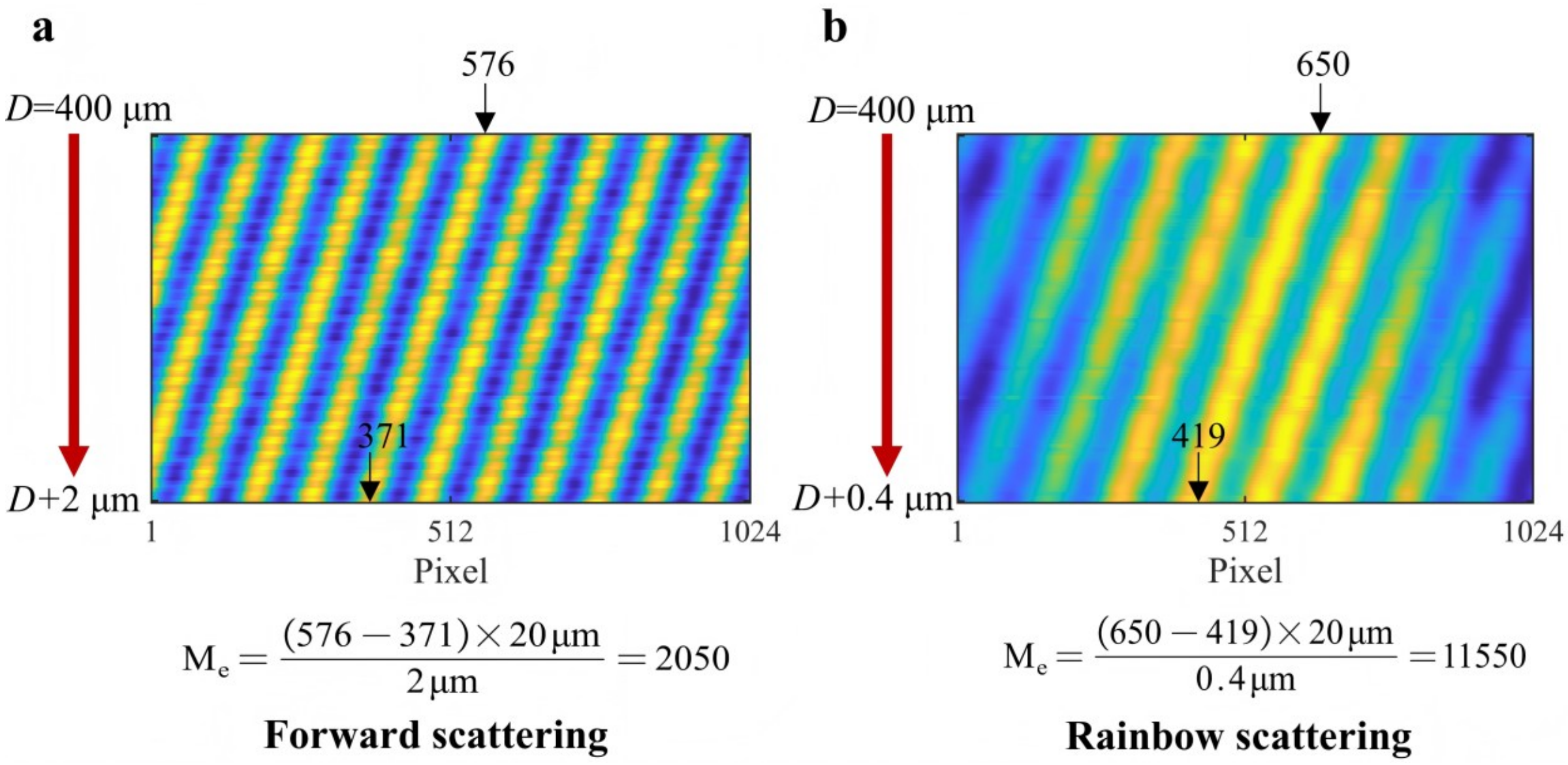

**Fig. S2** Equivalent magnification of fringe displacement caused by jet diameter variation. Assuming a 1° scattering angle is imaged onto 1024 pixels (pixel size 20 μm), the displacement of an extremum is converted to a camera-plane shift. **a,** Forward scattering: a 2 μm diameter change gives a 4100 μm shift, yielding an equivalent magnification of 2050. **b,** Rainbow scattering, a 0.4 μm diameter change gives a 6420 μm shift, yielding an equivalent magnification of 11550.

## Noise from the syringe pump

In our early experiments, the jets were generated using a syringe pump, which introduced significant noise into the initial disturbance measurements at $z$ = 2.5, as shown in Fig. S3. The corresponding Fast Fourier Transform (FFT) result reveals substantial low-frequency components.

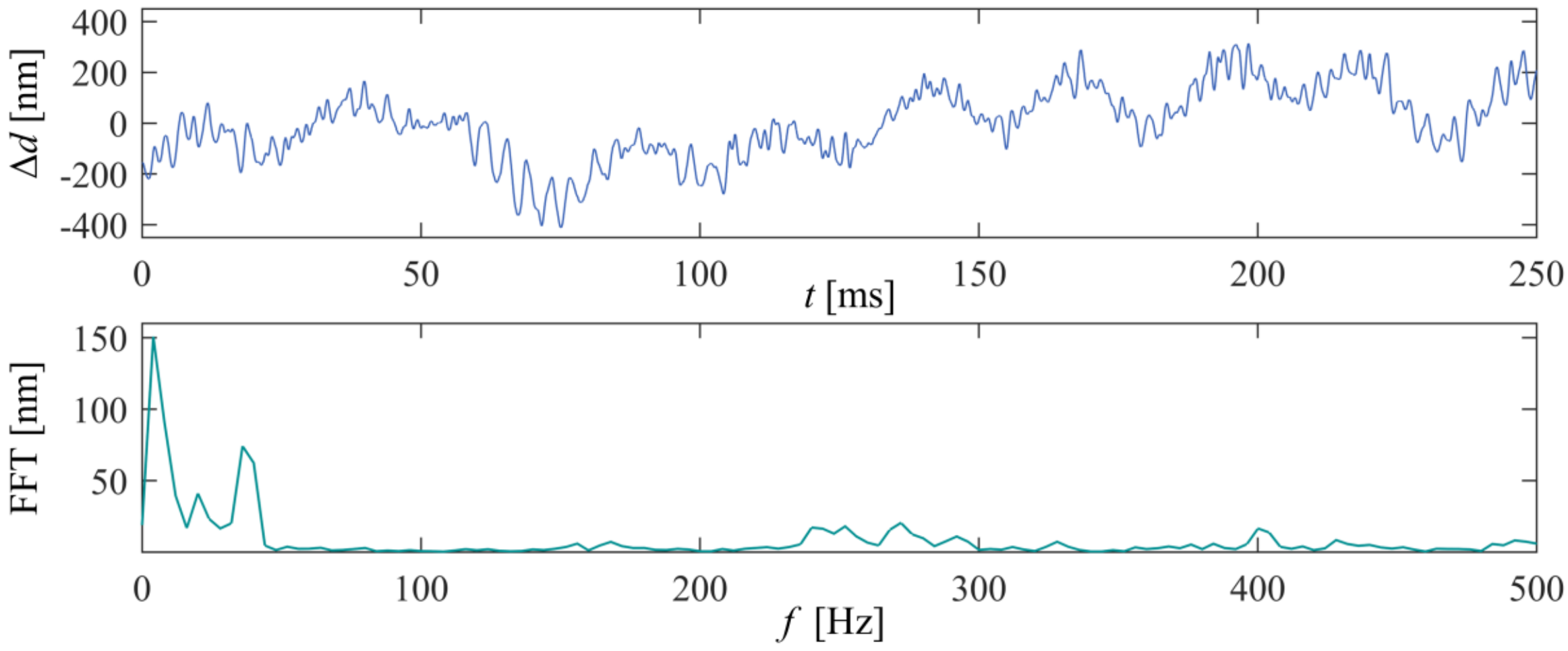

**Fig. S3** Background noises in the initial disturbance measurement at $z$ = 2.5 from syringe pump.

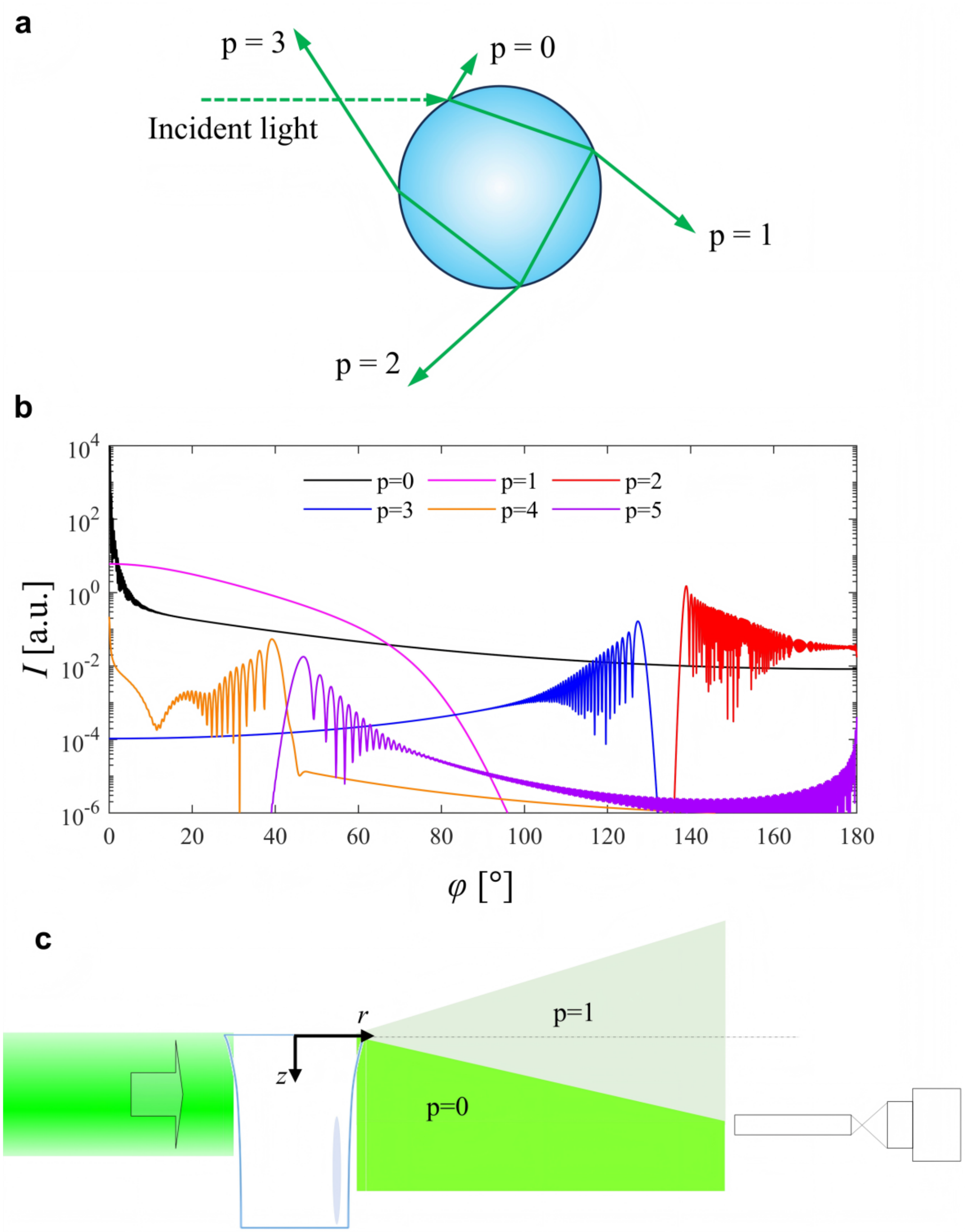

**Fig. S4** Decomposition of scattered light by ray order and the effect of jet contraction. **a,** Schematic of light rays with different numbers of internal refractions, labeled by p. **b,** Angular intensity distributions for different p, computed from the Debye series expansion of Lorenz-Mie theory [1]. The total intensity is the sum over all p. **c,** Separation of the p=0 and p=1 rays due to the jet contraction at the nozzle exit, showing that the expected dark and bright fringes appear only in the overlap region of the two rays.

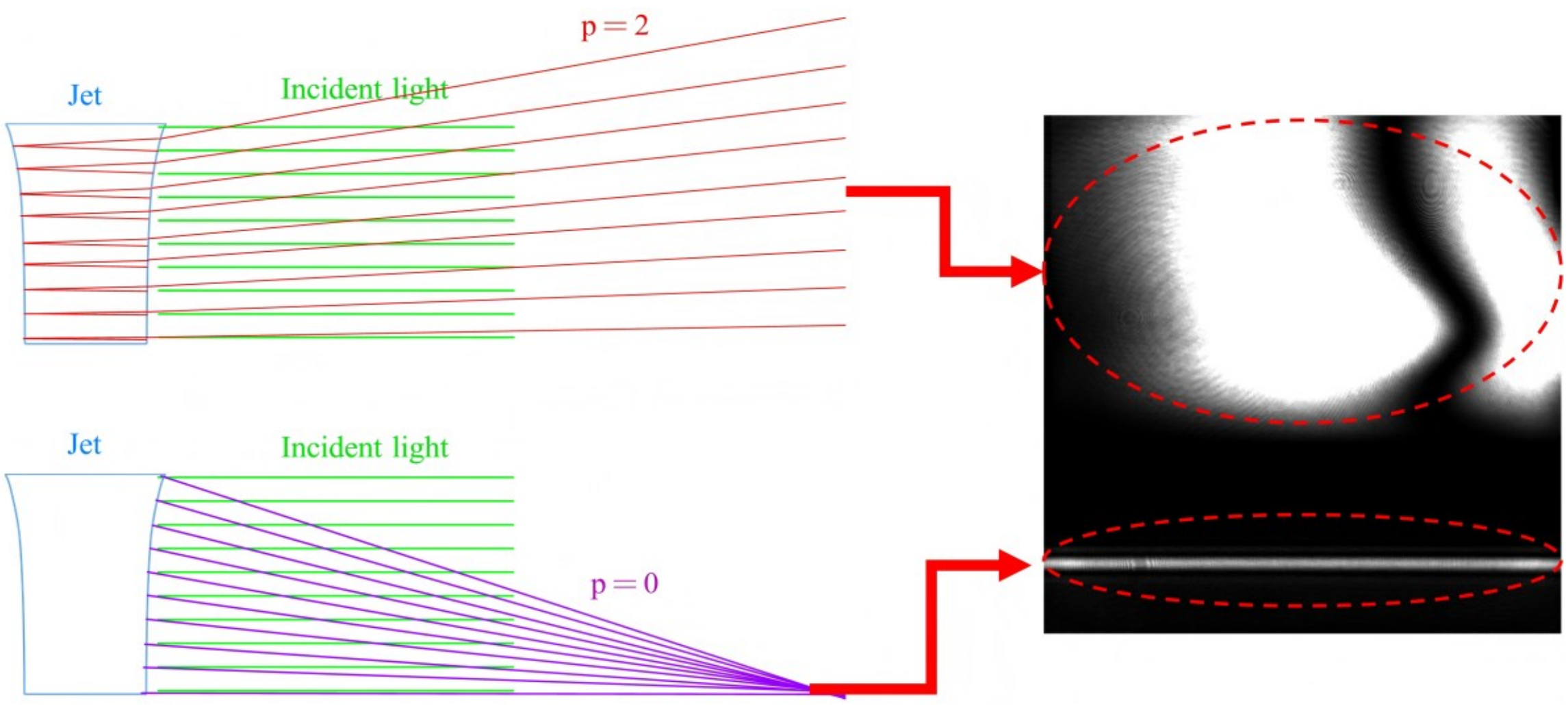

**Fig. S5** Geometric optics illustration of ray separation at the rainbow scattering region. The captured fringe image exhibits a broad fringe at the top (p=2) and the thin bright line at the bottom (p=0). The typical rainbow pattern predicted by Lorenz-Mie theory appears only when the light rays from the top and bottom paths coincide spatially.

**Determination of jet mean diameter**

The jet mean diameter $d_0$ serves as the starting point for the fringe-to-disturbance inversion. In a fixed optical setup, the density of the scattering fringes reflects the jet mean diameter. To determine $d_0$, we apply a calculation based on the same algorithm shown in Fig. S6, but with a much larger step size $\alpha$. An example is provided in Fig. S6. A water jet was generated from a 400 μm nozzle, and scattering fringes were captured at several locations with distinct fringe densities. By performing Steps 1 and 2 of Fig. 5, the cosine vector shown in Fig. S6a can be obtained. The result exhibits an oscillatory pattern with varying amplitude, where the amplitude represents the similarity between the experimental fringe density and the theoretical ones. Thus, by extracting the envelope of the data and locating its maximum, the jet mean diameter can be determined. Fig. S6b presents five jet images, with the laser spot indicating the scattering measurement point. The mean diameters derived from direct imaging and from the scattering method are compared in Fig. S6c and show good agreement.

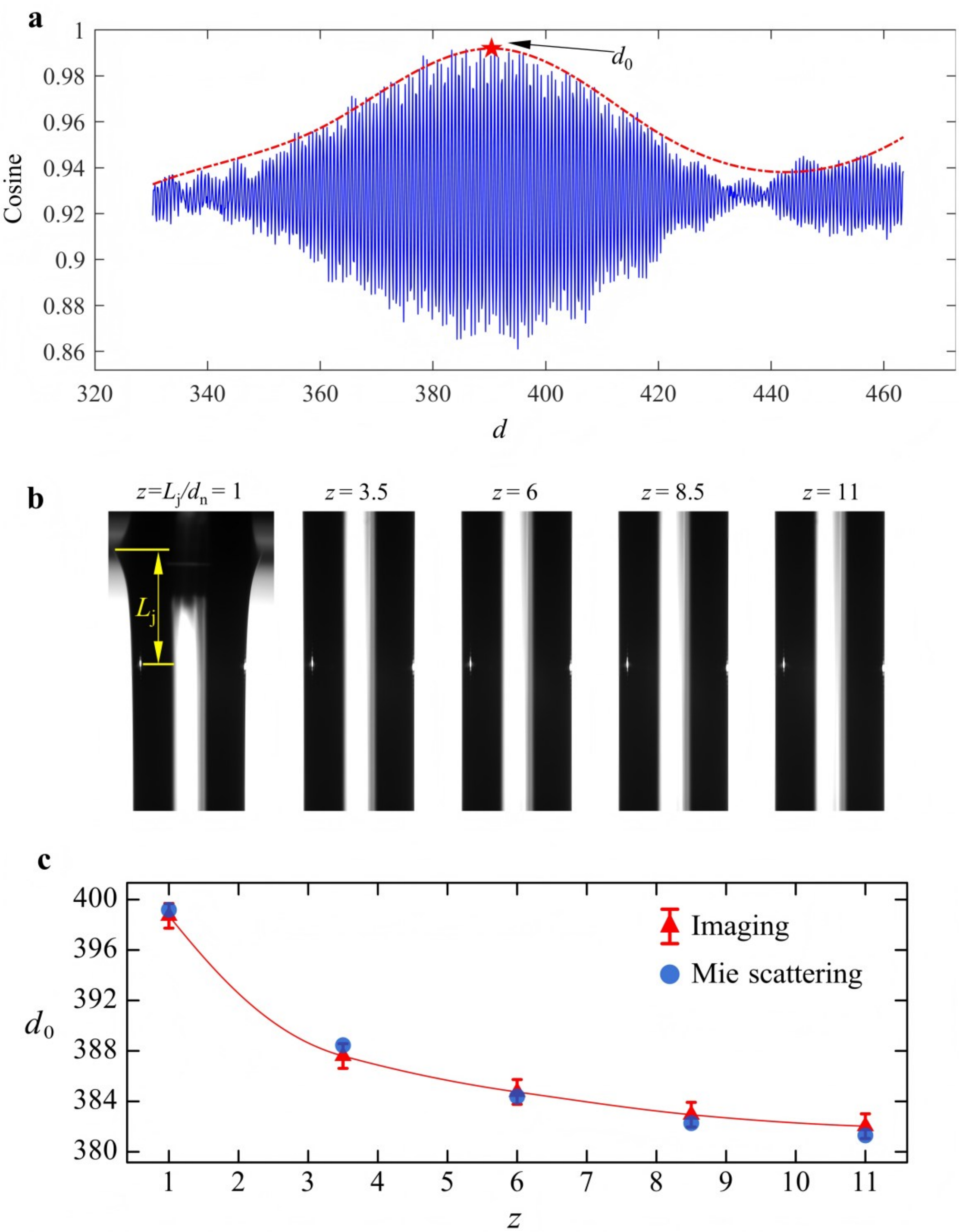

**Fig. S6 a**, Determination of the jet mean diameter using the envelope mehod. **b**, Jet images with the laser spot marking the measurement point. **c**, Comparison of mean diameters from imaging and scattering.

As can be seen in Fig. S6, a micrometer-level deviation exists in the jet mean diameter, indicating the limited resolution of the envelope method. However, the disturbances—i.e., the variations in jet diameter—are primarily reflected as phase shifts in the scattering fringes (Fig. S1). Fig. S7 examines how much this mean-diameter deviation affects the disturbance measurements and shows that its influence remains limited. This robustness is not surprising, since the rate of change of the scattering fringes remains consistent over a certain range of adjacent diameters. Therefore, the value of $d_0$ obtained from Fig. S6 is fully acceptable.

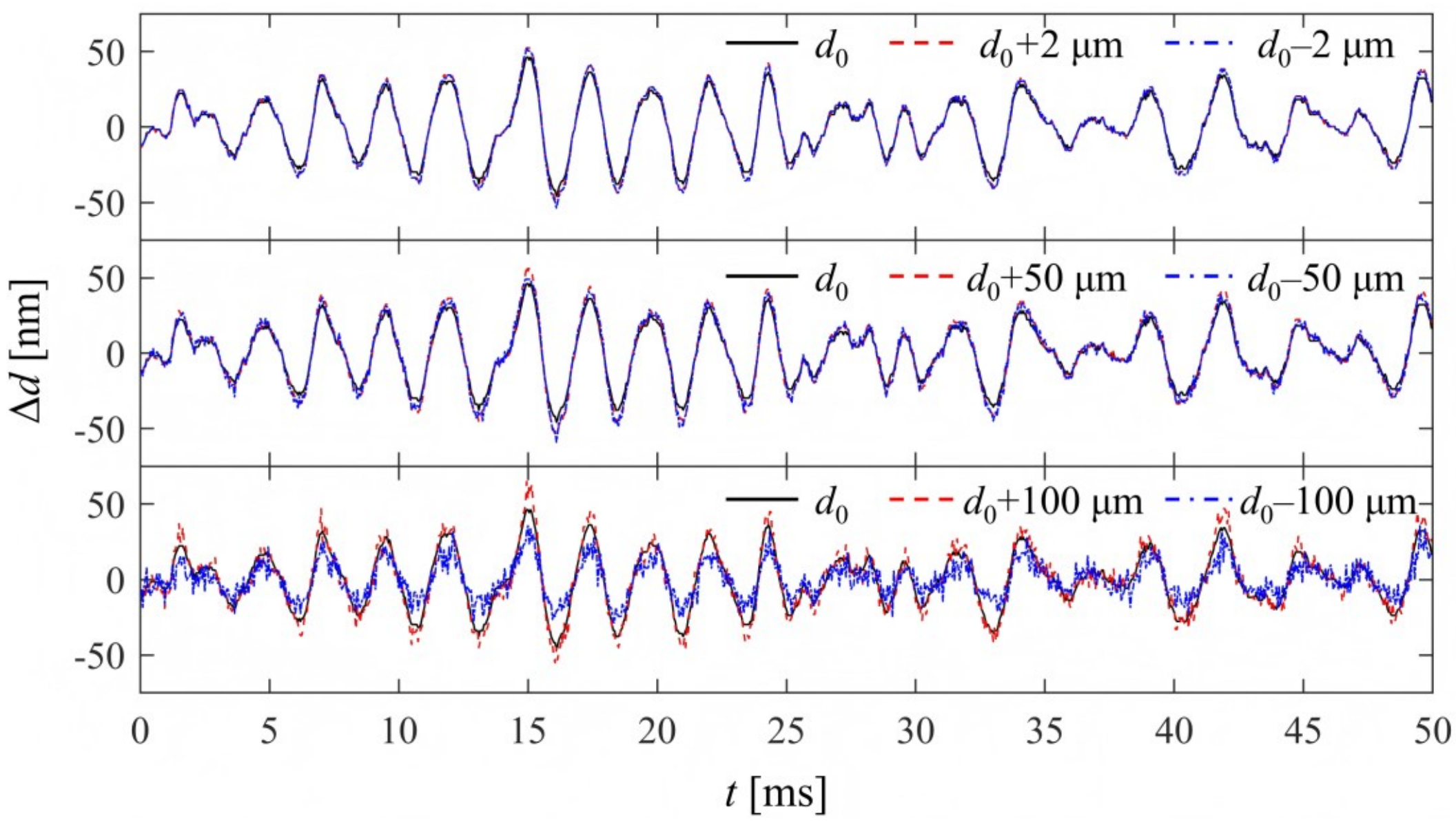

**Fig. S7** Assesment of the influence of mean-diameter deviation on the inverted disturbance measurements.

**Accuracy evaluation of the measurement method**

We first consider the propagation path of the measurement information. The variation in jet diameter is converted by light scattering into an angular distribution of light intensity. This intensity distribution is then transformed by an imaging system into discrete digital grayscale values by the camera sensor. Finally, the algorithm described in the main text (Fig. 9) reconstructs the diameter variation from the grayscale changes. The error propagation chain is illustrated in Fig. S8.

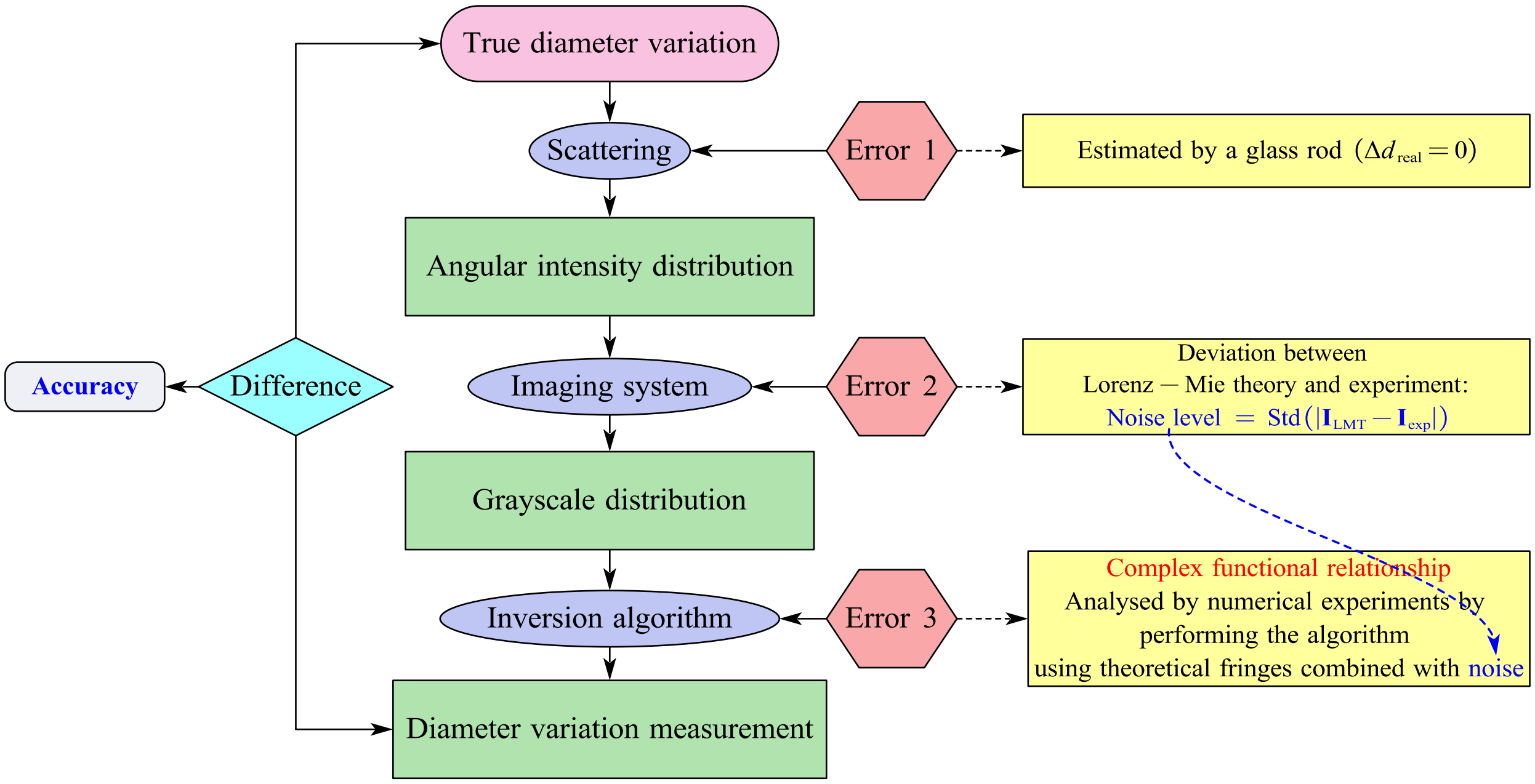

**Fig. S8** Schematic of the error propagation path in the measurement

The true diameter variation is mapped to the angular distribution of scattered light through the scattering process, which may introduce errors. First, it is essential to comfirm that the diameter change indeed causes a change in the intensity distribution. Otherwise, the entire measurement would be invalid. This physical basis is considered reliable, so the scattering process itself does not introduce a systematic error. Instead, errors in this step stem from deviations between the actual experiment and ideal conditions, such as laser wavelength instability, air turbulence, and uncompensated vibrations. These factors are assessed collectively using a solid glass rod (with invariant diameter), as shown in Fig. 13 of the main text. Another potential error source in the scattering step is the deviation of the measurement point from an ideal cylindrical shape (due to axial curvature), which has been discussed in Section 2.2.2.

The angular intensity distribution of the scattered light is converted into digital grayscale values by the imaging system, and this conversion introduce errors. Additionally, because the incident light is shaped into a laser sheet, the error associated with this process also manifests in the recorded grayscale images. This error level can be evaluated by comparing the experimental fringe pattern with the theoretical Lorenz-Mie calculation, specifically by computing the standard deviation of the absolute difference between the two signals:

$$\text{Error level} = \text{Std}\left(\left|I_{\text{LMT}} - I_{\text{exp}}\right|\right) \tag{S2}$$

Based on the data in Fig. 4 of the main text, this error level is estimated to be about 10 grayscale units. That is, the actual grayscale distribution can be modelled as the theoretical grayscale map (obtained by mapping the theoretical intensity to 0~255) plus a Gaussian noise with an amplitude of 10.

In the third step, the captured fringes are converted into diameter variation via our inversion algorithm. Because the relationship between the scattered intensity distribution and the jet diameter is complex, the inversion process itself can introduce errors. In practice, the grayscale error from the previous step propagates through the inversion and significantly affects the measurement accuracy.

Based on the above analysis, we establish the following accuracy evaluation procedure. First, using the experimental parameters and given jet diameters, we compute theoretical intensity distributions corresponding to the diameters and map them to a 0~255 grayscale array. Then, we add Gaussian noise to simulate real captured fringes. Assuming a sinusoidal diameter variation with a fixed amplitude $A_\text{s}$ and frequency $f_\text{s}$: $\tilde{d}(t) = d_0 + A_\text{s}\sin(2\pi f_\text{s} t)$, we feed the corresponding noisy grayscale distribution into the inversion algorithm to obtain the reconstructed signal, and compare it with the true signal. This allows us to asses the accuracy in the theoretical domain. By progressively reducing the amplitude of the simulated signal, we can also evaluate the resolution of the method. In particular, when we set $A_\text{s} = 0$, the inversion output represents the background noise floor of the measurement system. This computed noise floor can be directly compared with the experimental results obtained from the glass rod.

We first analyze the rainbow fringe configuration. We generate a set of theoretical rainbow fringes, map them to gray scale, and add Gaussian noise with amplitude 10. Then, we simulate a sinusoidal signal at $f_{\mathrm{s}} = 1000\,\mathrm{Hz}$ with sampling frequency $F_{\mathrm{s}} = 20{,}000\,\mathrm{Hz}$ ( $t_i = i/F_{\mathrm{s}}$) and various amplitudes $A_{\mathrm{s}}$. The corresponding noisy grayscale patterns are fed into the inversion algorithm to retrieve the diameter variation.

For $A_{\mathrm{s}} = 0$, the result is shown in **Fig. S9**a. Although the true amplitude is zero, the reconstructed signal exhibits oscillations, most likely due to the propagation of the added Gaussian noise. The RMS (root mean square) of the reconstructed signal is 0.17 nm, which agrees well with the glass-rod experimental result (Fig. 13). The power spectral density (PSD) from this numerical experiment also matches the experimental PSD from the glass rod. This agreement not only validates the numerical approach that inversion from noise-added theoretical fringes is credible, but also confirms that the primary source of inaccuracy in our method is the deviation between the experimental fringes and the ideal ones, which propagates through the inversion to the final result.

Next, we set $A_{\mathrm{s}} = 0.1$ nm and perform the same procedure, and the results is shown in Fig. S9b. In the time domain, the reconstructed signal appears to deviate significantly from the true signal, However, after FFT, the algorithm identifies the dominant frequency, and the amplitude at that frequency differs from the true value by only 7%. All other frequency components are below 0.025 nm.

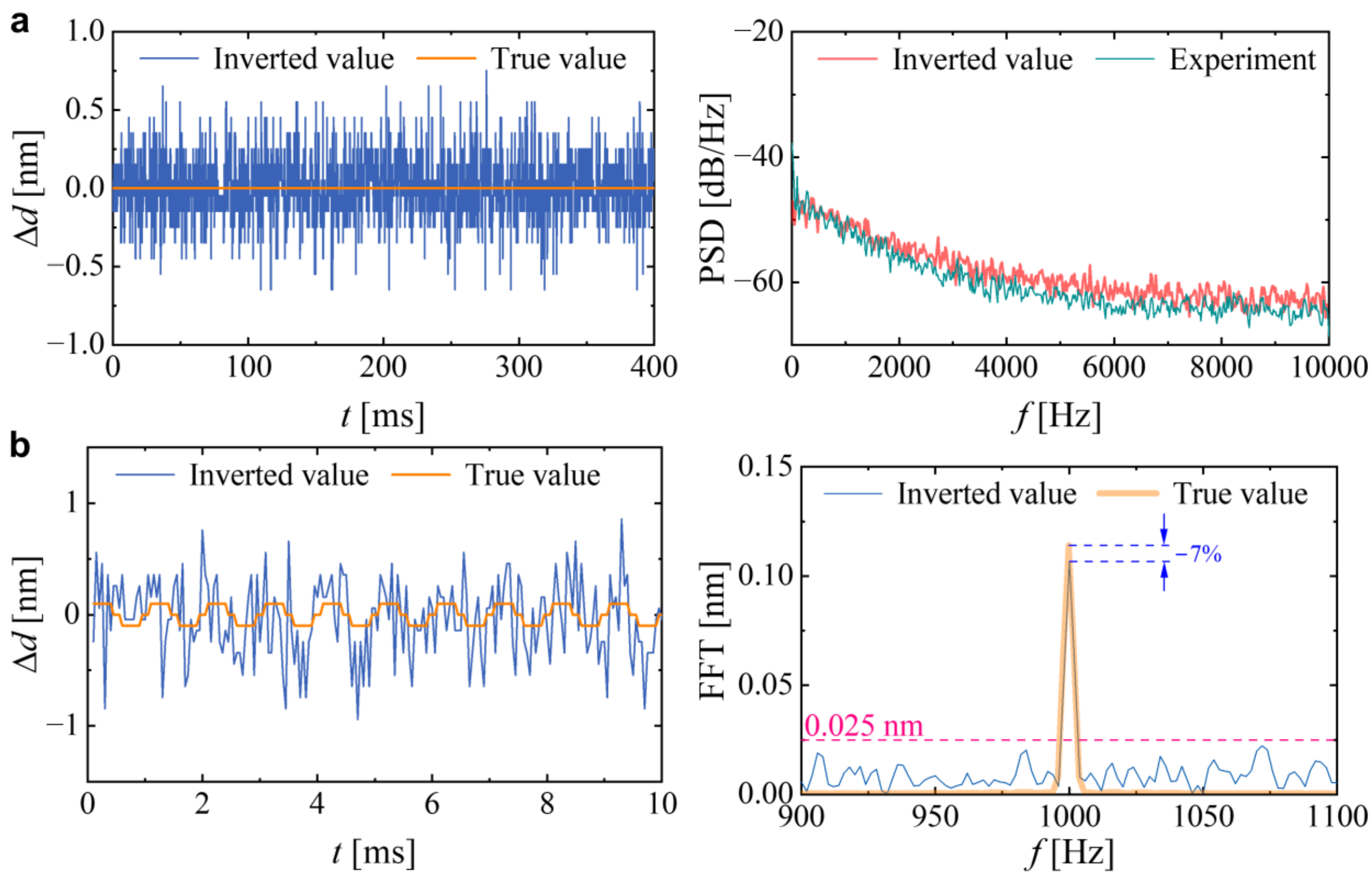

**Fig. S9** Reconstructed signals by the numerical simulation in the rainbow-scattering configuration. **a**, $A_s = 0$. **b**, $A_s = 0.1$ nm.

We then repeat the test for $A_s = 0.5, 1, 5$ nm (**Fig. S10**). As the amplitude increases, the time-domain reconstruction becomes increasingly close to the truth, and the FFT amplitudes at the dominant frequency also show excellent agreement. These results demonstrate that the proposed inversion algorithm is highly capable of capturing subtle fringe variations. The residual deviation is primarily due to the fringe error. Future improvements in the optical setup and image acquisition to obtain lower-noise fringes would further enhance the measurement accuracy.

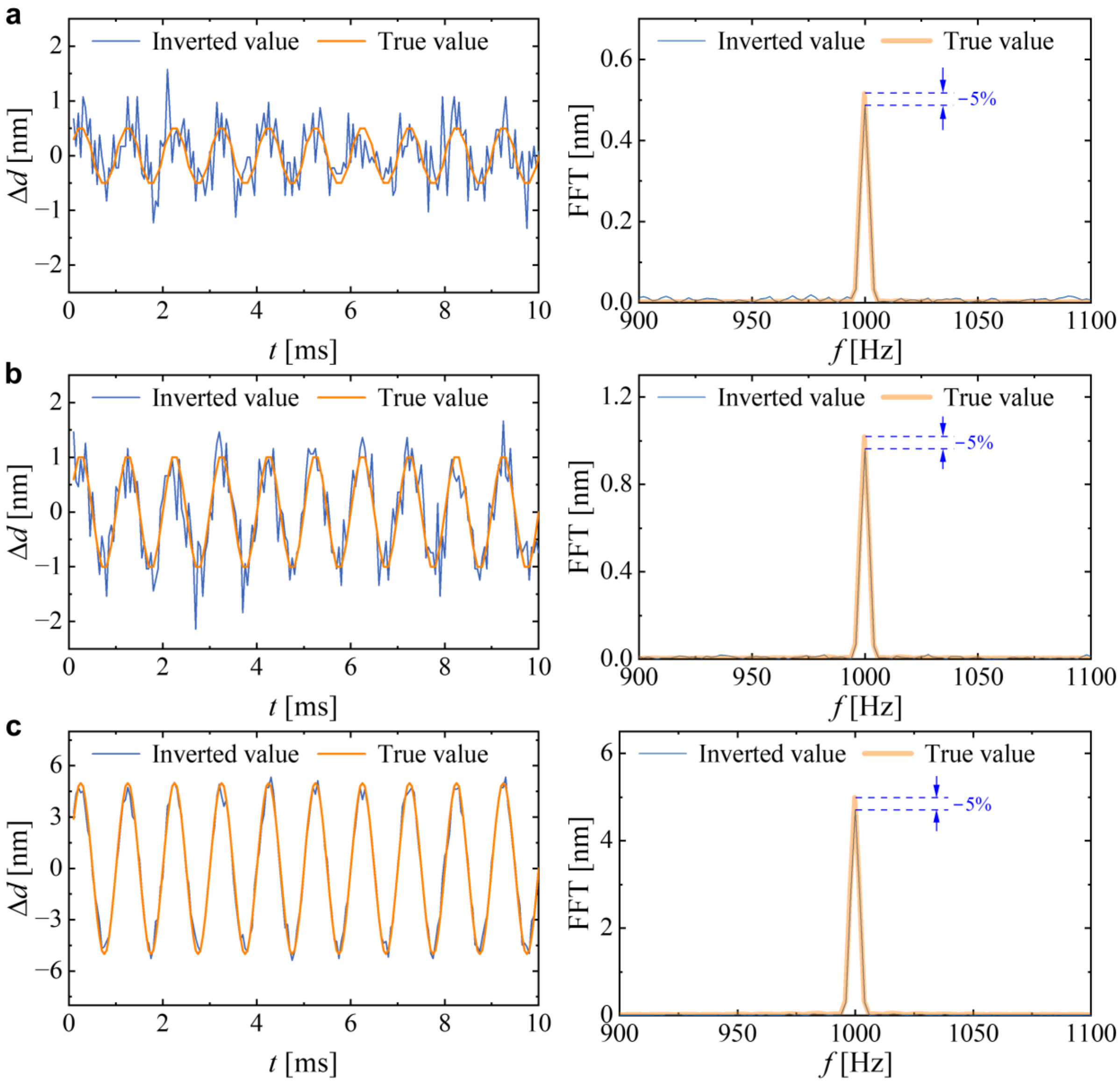

**Fig. S10** Reconstructed signals and their FFTs by the numerical simulation in the rainbow-scattering configuration. **a**, $A_s = 0.5$ nm. **b**, $A_s = 1$ nm. **c**, $A_s = 5$ nm.

All the preceding discussions are based on rainbow fringes. The forward-scattering configuration differs slightly because its sensitivity is lower than that of rainbow scattering. **Fig. S11** presents the inversion results for forward-scattering fringes with $A_s = 0$ and $0.1$ nm. For $A_s = 0$ (**Fig. S11**a), the RMS is 0.33 nm, and the PSD level is slightly higher than that from the glass-rod experiment. Nevertheless, for $A_s = 0.1$ nm (**Fig. S11**b). our algorithm still correctly recovers the dominant frequency in the frequency domain, although the amplitude discrepancy is somewhat larger. All other frequency components are below 0.05 nm.

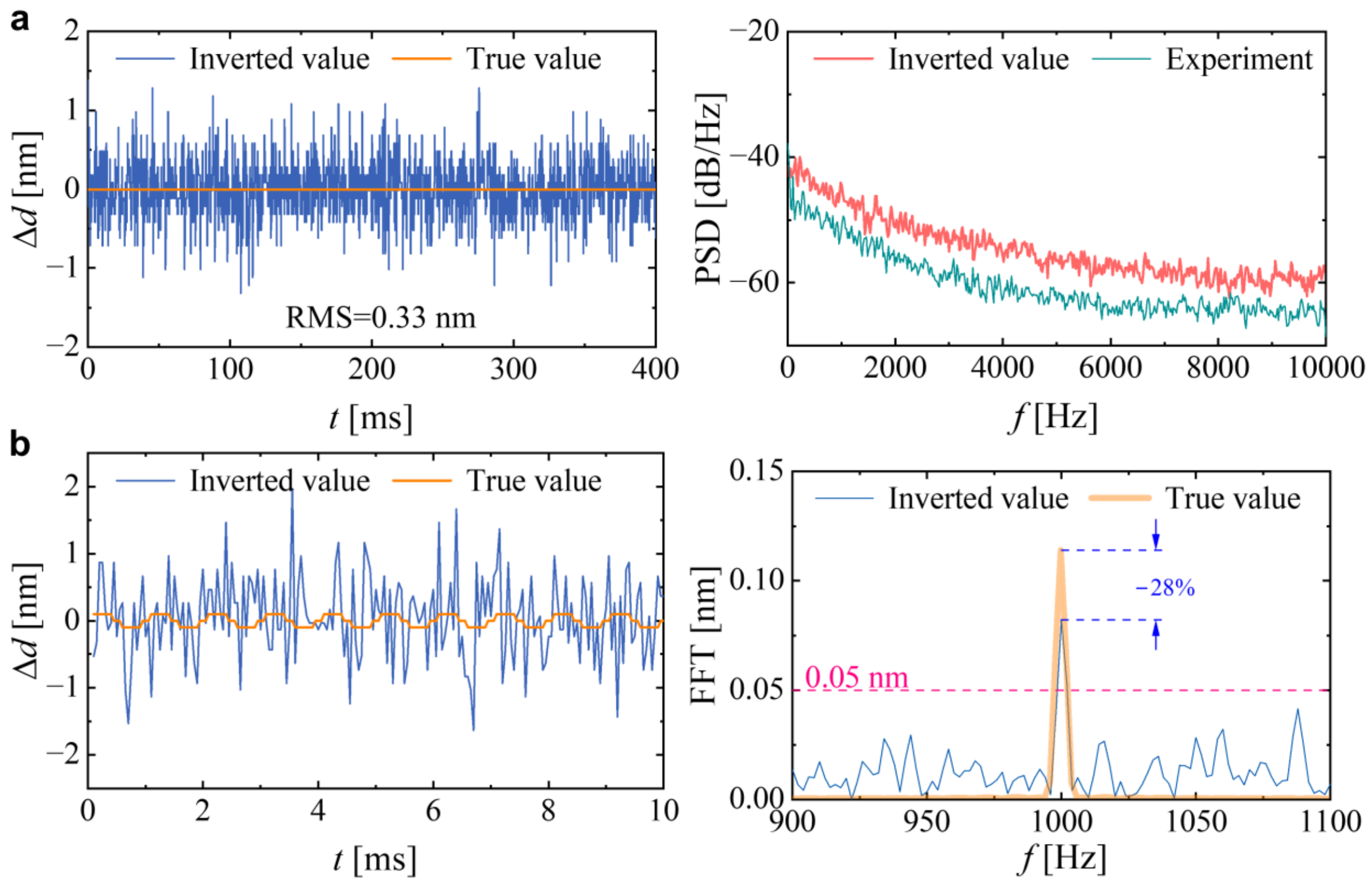

**Fig. S11** Reconstructed signals by the numerical simulation in the forward-scattering configuration. **a**, $A_s = 0$. **b**, $A_s = 0.1$ nm.

We further test $A_s = 4, 40$, and $400$ nm (**Fig. S10**). As the simulated diameter variation amplitude increases, the reconstructed signal from the noisy forward-scattering fringes approaches the true value more closely, because the noise becomes relatively less significant. Although the accuracy at the nanometer scale is somewhat inferior to that of the rainbow configuration, at the scale of hundreds of nanometers the reconstruction is nearly identical to the truth. Therefore, we adopt the following strategy: for amplitudes below 10 nm, we use rainbow-scattering fringes; for amplitudes above 10 nm, we employ forward-scattering fringes. The reason for not using rainbow-scattering fringes exclusively is that their very high sensitivity causes excessively drastic fringe changes when the diameter amplitude becomes large, which exacerbates the issue discussed in Section 2.3.3 of the main text.

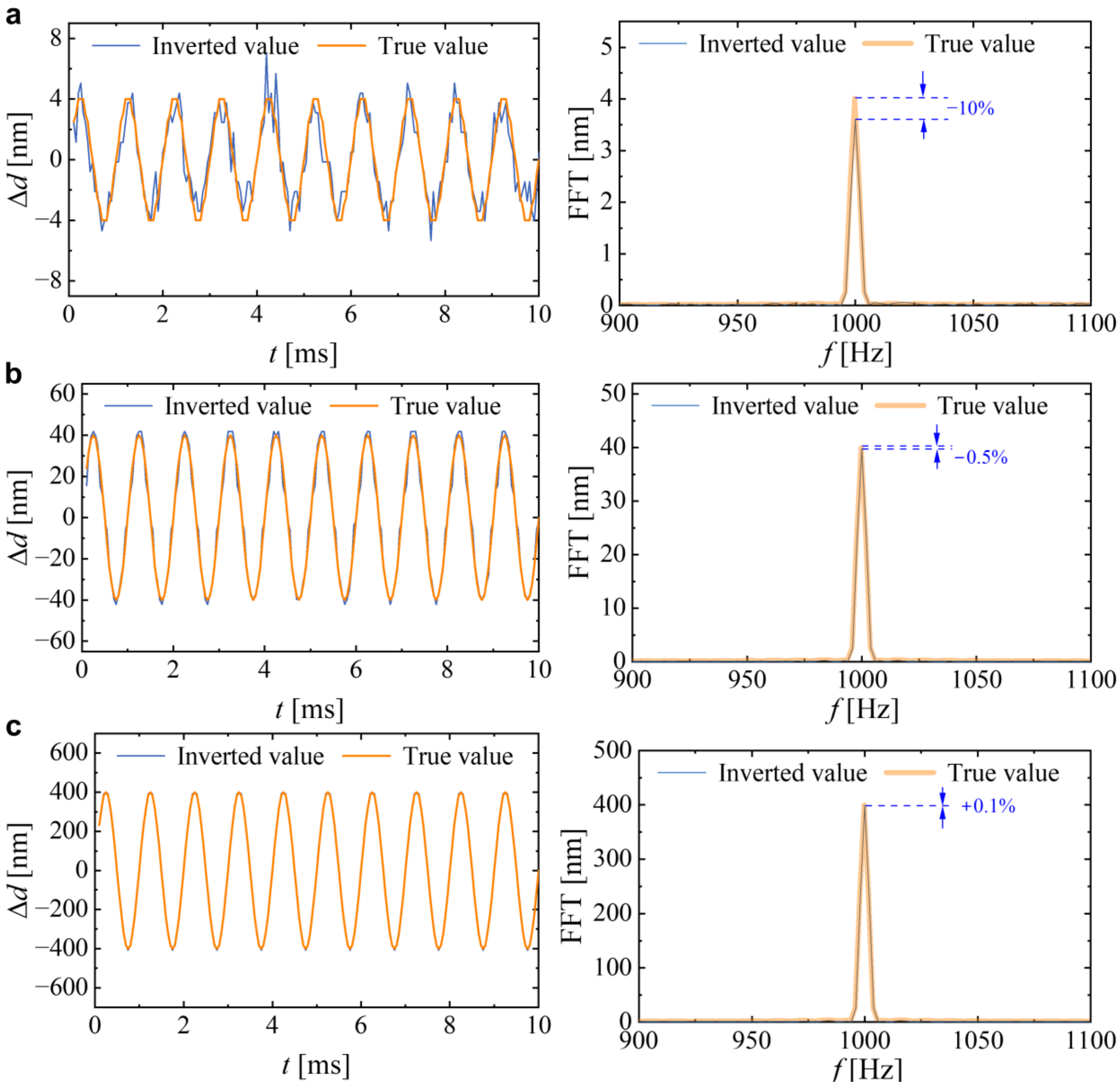

**Fig. S12** Reconstructed signals and their FFTs by the numerical simulation in the forward-scattering configuration. **a**, $A_s = 4$ nm. **b**, $A_s = 40$ nm. **c**, $A_s = 400$ nm.

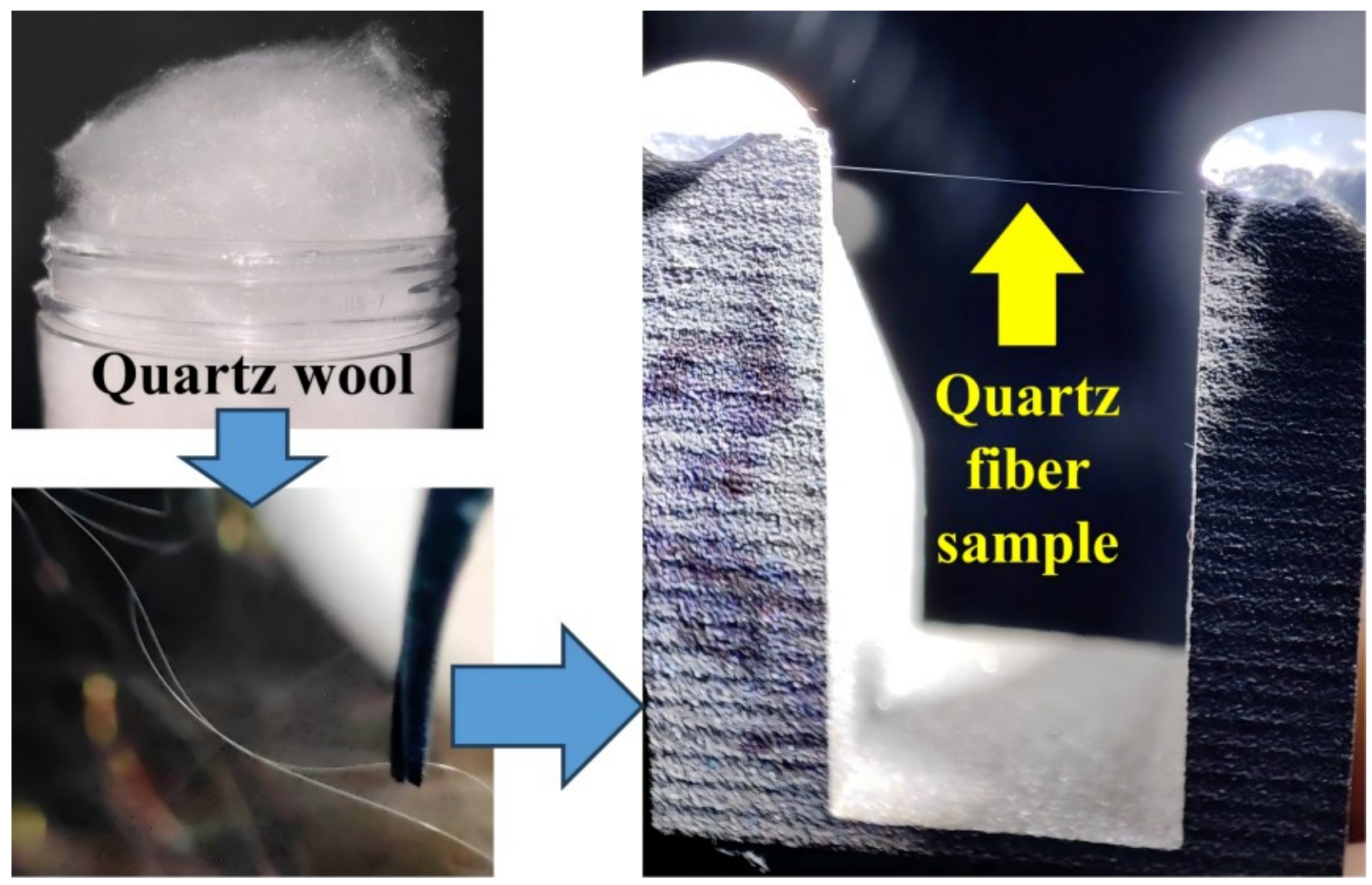

**Fig. S13** Preparation of the quartz fiber sample. A single quartz fiber extracted from quartz wool was secured onto a metal frame.

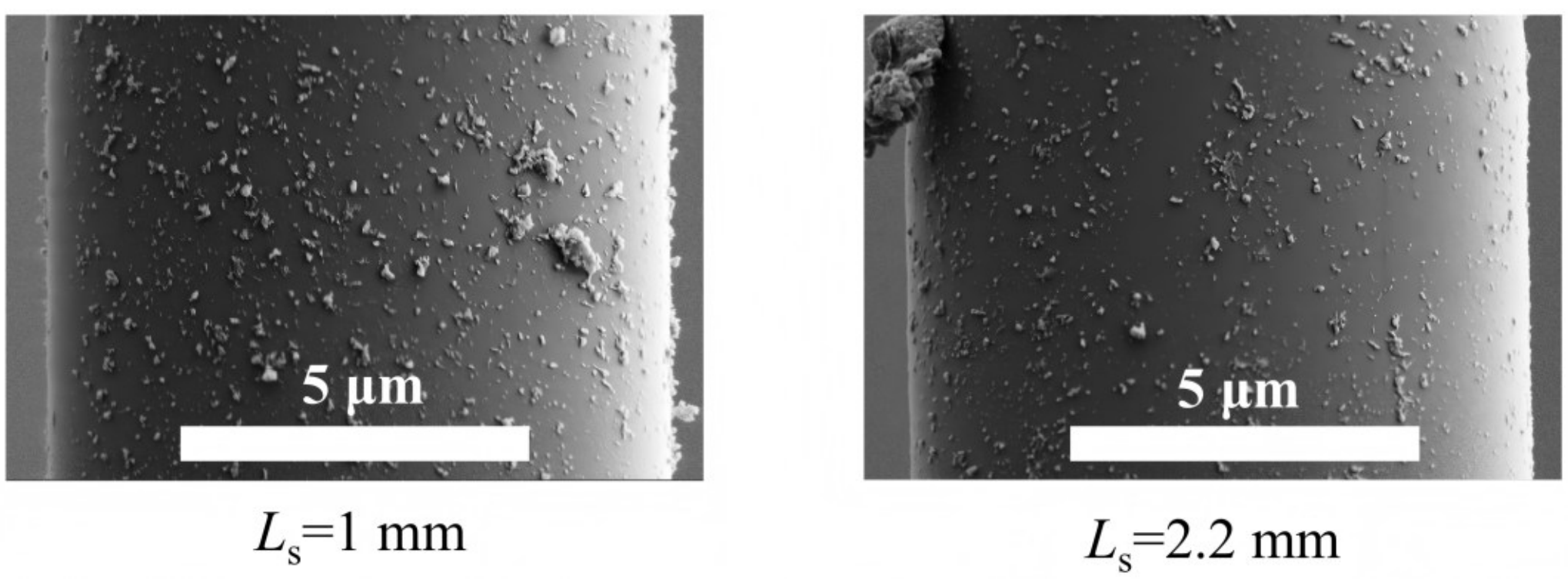

**Fig. S14** SEM images of the two points with the largest error margins in Fig. 12, showing the surface impurities responsible for the deviation.

## Calculation of the dominant frequency $f_d$

The spatial linear stability analysis based on a one-dimensional model [2] yields the dispersion relation:

$$\frac{1}{2}k^4 - 3iOh\sqrt{We}k^3 + \left(-\frac{1}{2} - We + 3iOh\omega\right)k^2 + 2\omega\sqrt{We}k - \omega^2 = 0, \qquad \text{(S3)}$$

where $k$ is the disturbance wave number, $\omega = 2\pi f$ with $f$ the disturbance frequency. By substituing $Oh = 0.0075$ and various $We$ numbers into the equation and solving for the imaginary part $k_\mathrm{i}$, the dispersion curves shown in Fig. S15 are obtained. The dominant frequency corresoponds to the maximum of $k_\mathrm{i}$.

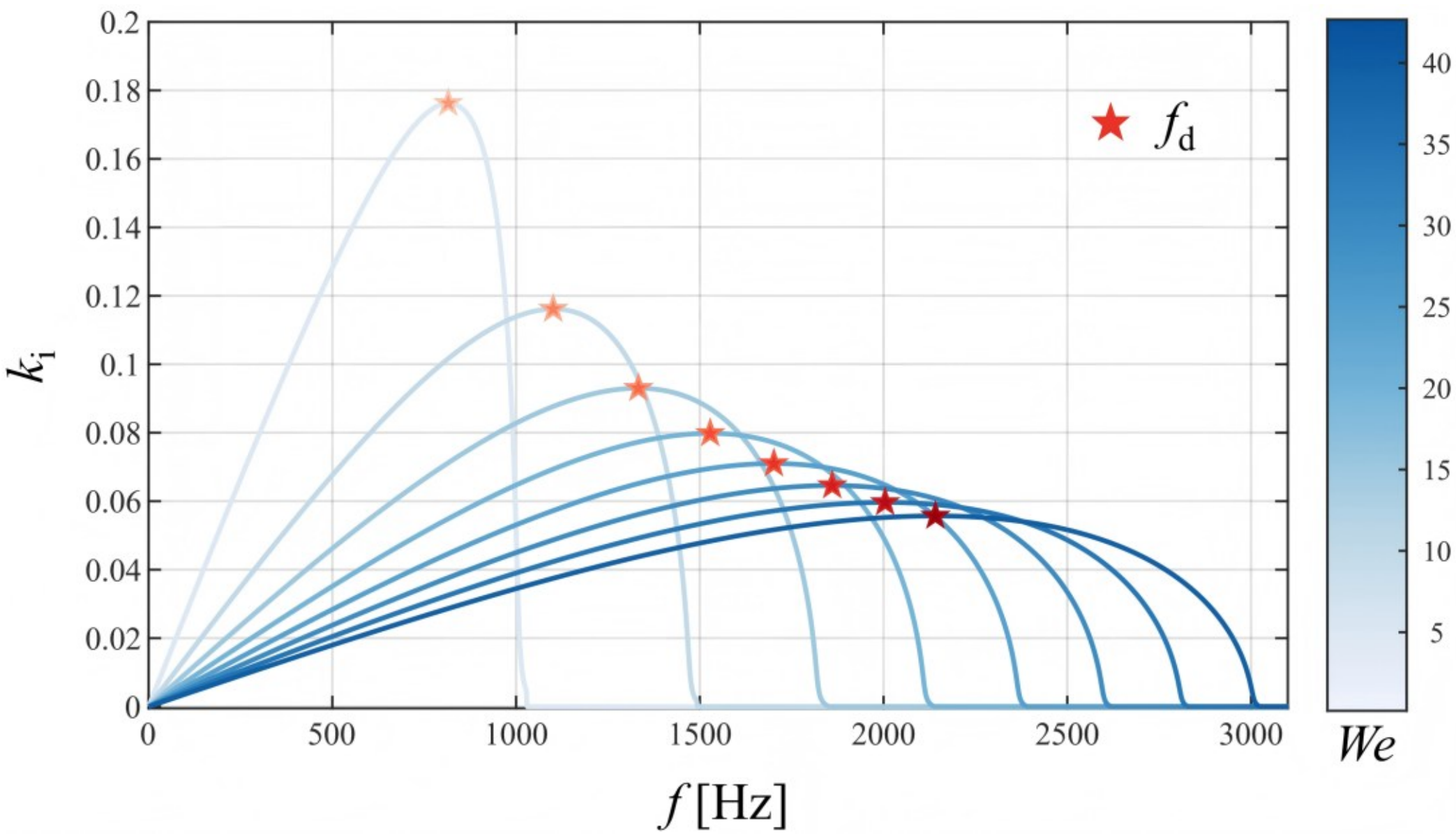

**Fig. S15** Dispersion curves obtained from the spatial linear stability analysis based on a one-dimensional model. The dominant frequencies are highlighted.

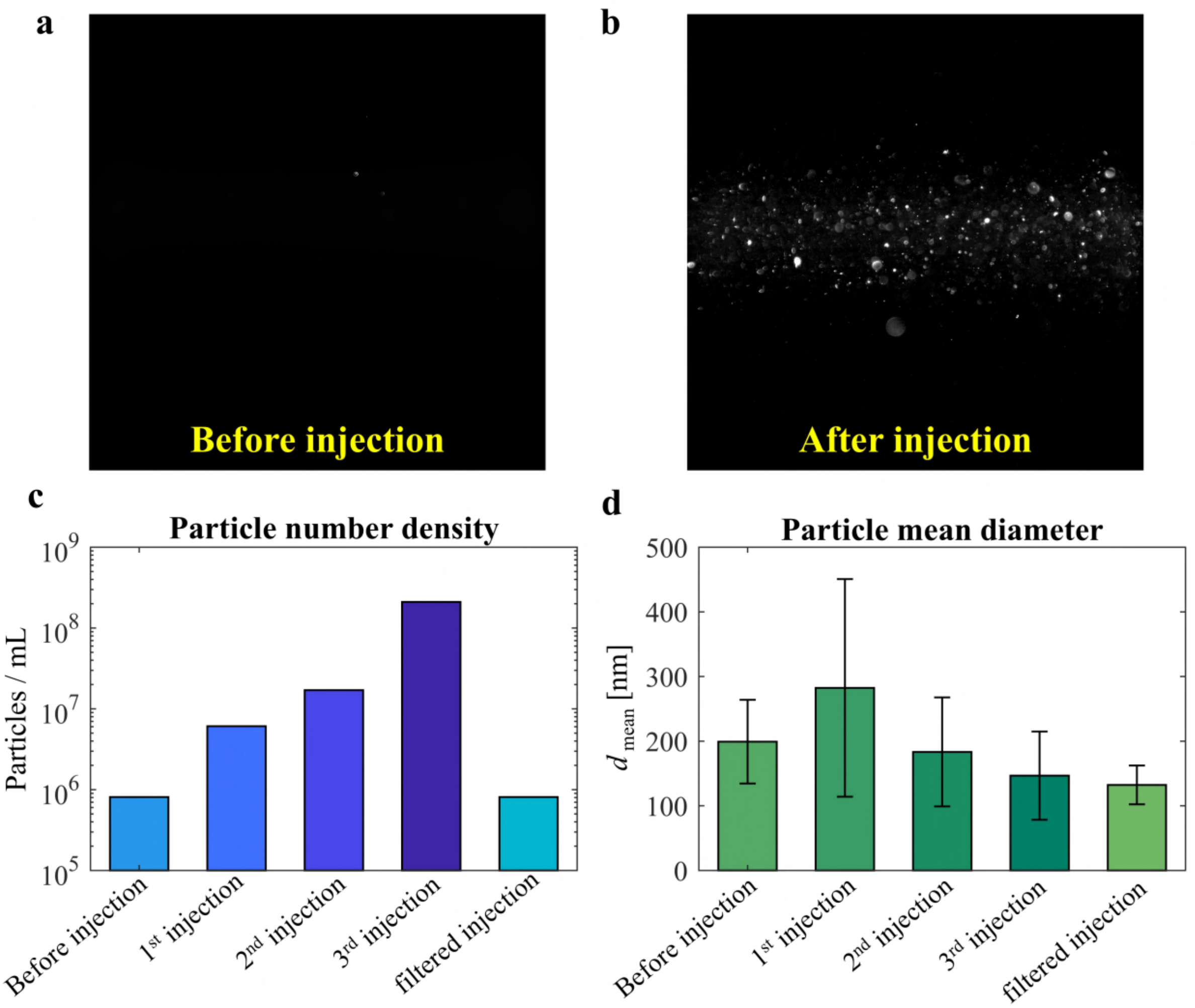

**Fig. S16** Formation of nanobubbles during water injection and their characterization by laser illumination and nanoparticle tracking analysis. **a,** Laser-illuminated image of bulk water before injection, showing almost no visible bubbles. **b,** Numerous bubbles appear after a single injection. **c,** Particle number density measured by Zeta View MONO, showing an increase with repeated injections and a return to pre-injection level after filtration through a PTFE membrane. **d,** Bubble size measured by ZetaView MONO, indicating a mean diameter in the range of 100-300 nm.

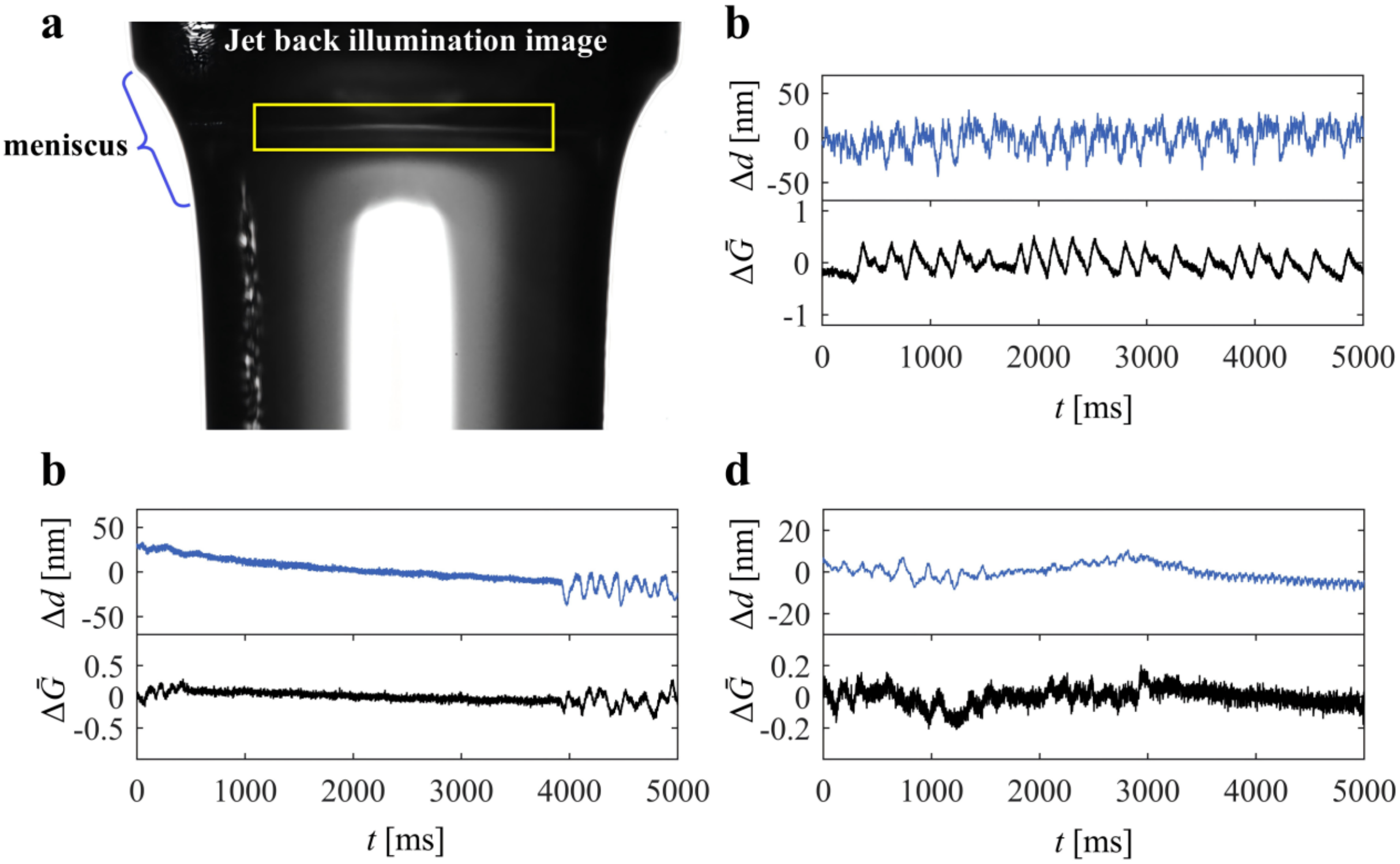

**Fig. S17** Relationship between meniscus dynamics and initial jet disturbances. a, Back-illuminated meniscus at the nozzle exit, where the meniscus geometry produces a distinct bright horizontal line (highlighted by the square). b-d, Synchronous measurements of the disturbance at $z$=2.5 and the mean grayscale oscillation $\Delta\bar{G}$ within the square-marked area for three independent experiments. The coincidence of low-frequency oscillations in both signals reveals the underlying link between internal meniscus instabilities and the initial disturbances.